%% file: main.tex
\documentclass[onecolumn,unpublished,a4paper]{quantumarticle}
\usepackage[utf8]{inputenc}
\usepackage[backend=biber,style=alphabetic,maxbibnames=999,backref=true]{biblatex}
\usepackage{amsmath}
\usepackage{amssymb}
\usepackage{amsthm}
\usepackage{amsfonts}
\usepackage{changepage}
\usepackage[caption=false]{subfig}
\usepackage[colorlinks]{hyperref}
\usepackage[all]{hypcap}
\usepackage{import}
\usepackage{tikz}
\usepackage{relsize}
\usepackage{color,soul}
\usepackage{capt-of}
\usepackage{mathtools}
\usepackage{float}
\usepackage[section]{placeins}
\usepackage{mdframed}
\usepackage{listings}
\usepackage[T1]{fontenc}  
\usetikzlibrary{decorations.pathreplacing}
\usepackage{braket}

\DefineBibliographyStrings{english}{
  backrefpage  = {Cited on page},
  backrefpages = {Cited on pages}
}
\renewcommand\thesection{\arabic{section}}

\theoremstyle{definition}

\theoremstyle{definition}

\theoremstyle{definition}

\newcommand{\eq}[1]{\hyperref[eq:#1]{Equation~\ref*{eq:#1}}}
\renewcommand{\sec}[1]{\hyperref[sec:#1]{Section~\ref*{sec:#1}}}
\DeclareRobustCommand{\app}[1]{\hyperref[app:#1]{Appendix~\ref*{app:#1}}}
\newcommand{\fig}[1]{\hyperref[fig:#1]{Figure~\ref*{fig:#1}}}
\newcommand{\tbl}[1]{\hyperref[tbl:#1]{Table~\ref*{tbl:#1}}}
\newcommand{\theoremref}[1]{\hyperref[theorem:#1]{Theorem~\ref*{theorem:#1}}}
\newcommand{\definitionref}[1]{\hyperref[definition:#1]{Definition~\ref*{definition:#1}}}

\newcommand{\rangesum}[2]{\sum_{#1=0}^{#2-1}}
\newcommand{\rangeprod}[2]{\prod_{#1=0}^{#2-1}}
\newcommand{\len}[0]{\text{len }}
\newcommand{\negif}[0]{\text{negif}}

\DeclareFixedFont{\ttb}{T1}{txtt}{bx}{n}{8}
\DeclareFixedFont{\ttm}{T1}{txtt}{m}{n}{8}
\definecolor{deepblue}{rgb}{0,0,0.5}
\definecolor{deepred}{rgb}{0.6,0,0}
\definecolor{deepgreen}{rgb}{0,0.5,0}
\newcommand{\pythonlisting}[2]{\lstinputlisting[
    language=Python,
    basicstyle=\fontsize{#1}{#1}\ttfamily,
    commentstyle=\fontsize{#1}{#1}\ttfamily\bfseries\color{deepgreen},
    keywordstyle=\fontsize{#1}{#1}\ttfamily\bfseries\color{deepblue},
    xleftmargin=0.5cm,
    frame=tlbr,
    framesep=0.2cm,
    framerule=1pt,
    morekeywords={self,None,assert,venting\_into,venting\_into\_new\_table,alloc\_quint,quint,z,del\_measure\_x,alloc\_phase\_gradient,qpu,QPU,del\_phase\_gradient,del\_by\_equal\_to\_const,push\_uncompute\_info,pop\_uncompute\_info,ghz\_lookup},
    keywordstyle={[4]\fontsize{#1}{#1}\ttfamily\color{deepred}},
    morekeywords={[4]Q\_residue,Q\_result,Q\_total,Q\_e,Q\_grad,Q\_target,Q\_helper,Q\_unresult,Q\_l1,Q\_l0,Q\_l,Q\_k,Q\_exponent,Q\_acc,Q\_dlog,Q\_out},
]{#2}}

\begin{document}
\title{
How to factor 2048 bit RSA integers with less than a million noisy qubits
}

\date{\today}
\author{Craig Gidney}
\email{craig.gidney@gmail.com}
\affiliation{Google Quantum AI, Santa Barbara, California 93117, USA}

\begin{abstract}
Planning the transition to quantum-safe cryptosystems requires understanding the cost of quantum attacks on vulnerable cryptosystems.
In Gidney+Eker{\aa} 2019, I co-published an estimate stating that 2048 bit RSA integers could be factored in eight hours by a quantum computer with 20 million noisy qubits.
In this paper, I substantially reduce the number of qubits required.
I estimate that a 2048 bit RSA integer could be factored in less than a week by a quantum computer with less than a million noisy qubits.
I make the same assumptions as in 2019: a square grid of qubits with nearest neighbor connections, a uniform gate error rate of $0.1\%$, a surface code cycle time of 1 microsecond, and a control system reaction time of $10$ microseconds.

The qubit count reduction comes mainly from using approximate residue arithmetic (Chevignard+Fouque+Schrottenloher 2024), from storing idle logical qubits with yoked surface codes (Gidney+Newman+Brooks+Jones 2023), and from allocating less space to magic state distillation by using magic state cultivation (Gidney+Shutty+Jones 2024).
The longer runtime is mainly due to performing more Toffoli gates and using fewer magic state factories compared to Gidney+Eker{\aa} 2019.
That said, I reduce the Toffoli count by over 100x compared to Chevignard+Fouque+Schrottenloher 2024.
\end{abstract}

\textbf{Data availability}: \emph{Code and assets created for this paper are available \href{https://doi.org/10.5281/zenodo.15347487}{on Zenodo}~\cite{gidneyzenodo2025factor}.}

{
  \renewcommand{\contentsname}{}
  \vspace{-1cm}

  \hypersetup{linkcolor=blue}
  \tableofcontents
}

\begin{figure}[ht]
    \centering
    \resizebox{\linewidth}{!}{
    \includegraphics{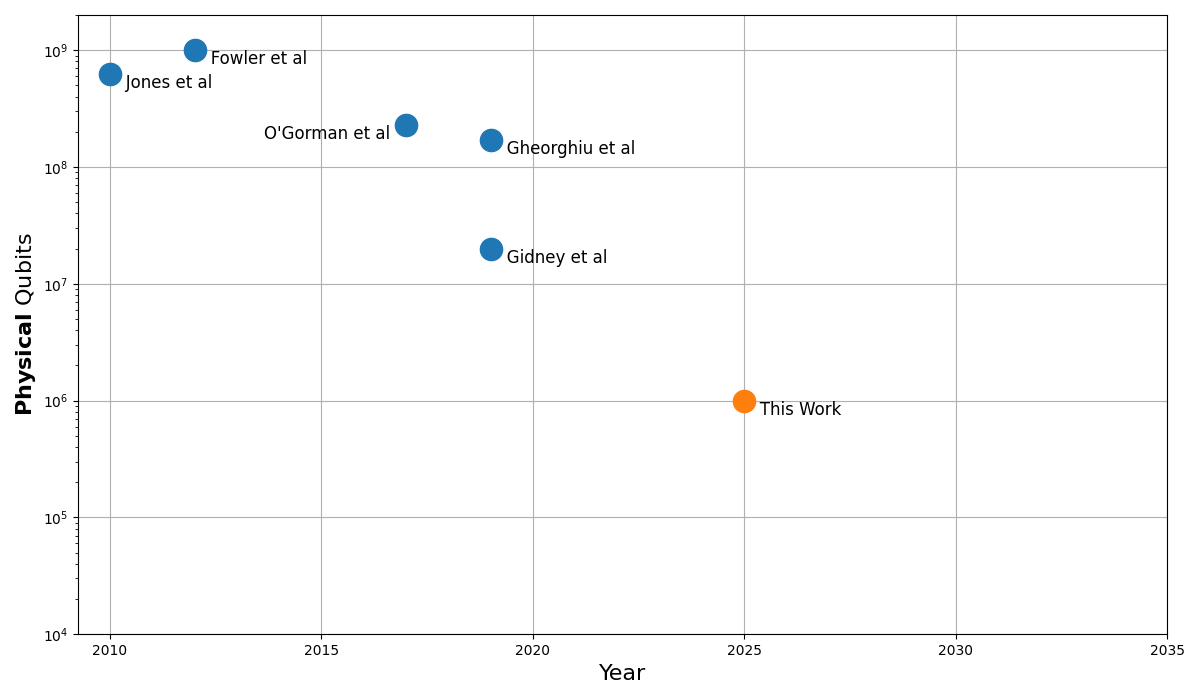}
    }
    \caption{
        Historical estimates, with comparable physical assumptions, of the physical qubit cost of factoring 2048 bit RSA integers.
        Includes overheads from fault tolerance, routing, and distillation.
        Results are from \cite{Jones2012,fowler2012surfacecodereview,OGorman2017,gheorghiu2019cryptanalysis,gidney2021factor}.
        Results such as \cite{VANMETER2010} and \cite{litinski2022activearchitecture} aren't included because they target substantially different assumptions or cost models.
    }
    \label{fig:historical-progression-physical}
\end{figure}

\begin{figure}[ht]
    \centering
    \resizebox{\linewidth}{!}{
    \includegraphics{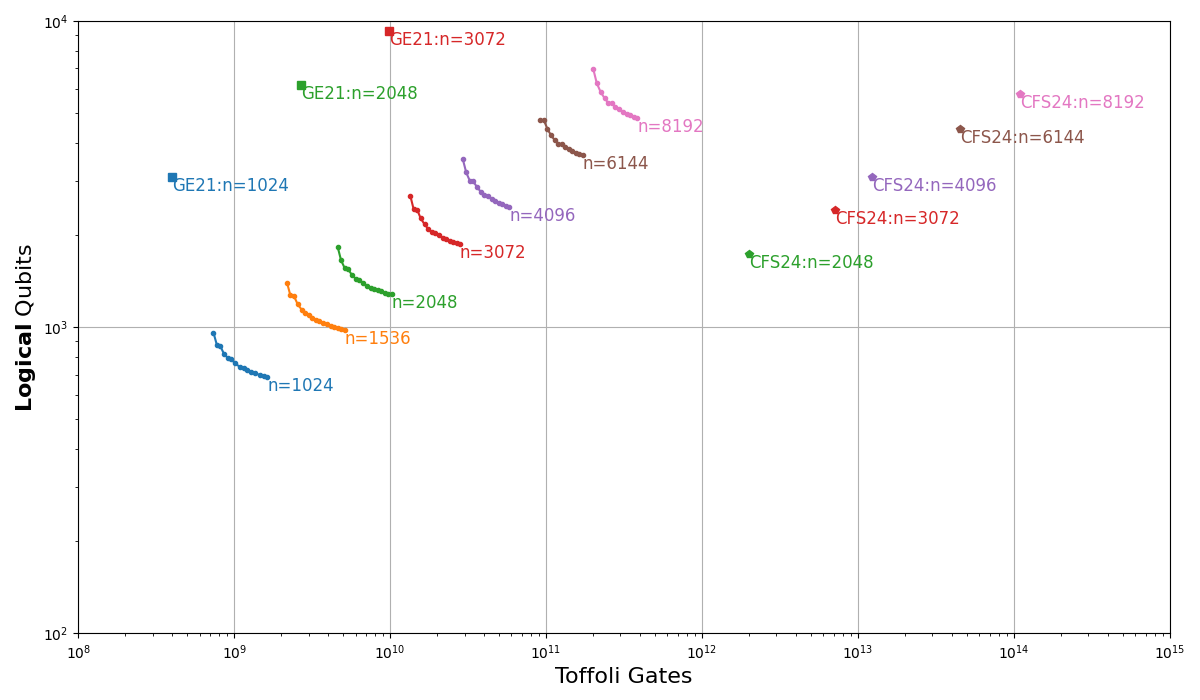}
    }
    \caption{
        Pareto frontiers achieved by this paper for the Toffoli and logical qubit cost of factoring $n$ bit RSA integers, for various values of $n$.
        This paper uses notably fewer logical qubits than \cite{gidney2021factor} (points labeled ``GE21'') and notably fewer Toffolis than \cite{chevignard2024reducing} (points labeled ``CFS24'').
    }
    \label{fig:pareto-curves}
\end{figure}

\section{Introduction}
\label{sec:introduction}

In 1994, Peter Shor published a paper showing quantum computers could efficiently factor integers~\cite{shor1994}, meaning the RSA cryptosystem~\cite{rivest1978} wasn't secure against quantum computers.
Understanding the cost of quantum factoring is important for planning and coordinating the transition away from RSA, and other cryptosystems vulnerable to quantum computers.
Correspondingly, since Shor's paper, substantial effort has gone into understanding the cost of quantum factoring~\cite{knill1995shor,Beckman1996,Vedral1996,zalka1998fast,Cleve2000,Beauregard2003,zalka2006pure,Whitney2009,VANMETER2010,fowler2012surfacecodereview,Jones2012,Pavlidis2014,ekeraa2016modifying,haner2016factoring,gidney2017factoring,OGorman2017,gheorghiu2019cryptanalysis,ekeraheuristic2019,ekera2021analyzeshortlogs,gidney2021factor,litinski2022activearchitecture,may2022compression,Regev2024,chevignard2024reducing} (and more).
A key metric is the number of qubits used by the algorithm, since this bounds the required size of quantum computer.

Historically, there was no known way to factor an $n$ bit number using fewer than $1.5n$ logical qubits~\cite{zalka2006pure,Beauregard2003,haner2016factoring,gidney2017factoring}.
Anecdotally, it was widely assumed that $n$ was the minimum possible because doing arithmetic modulo an $n$ bit number ``required'' an $n$ qubit register.
May and Schlieper had shown in 2019 that in principle only a single \emph{output} qubit was needed~\cite{may2022compression}, but there was no known way to prepare a relevant output that didn't involve intermediate values spanning $n$ qubits.
In 2024, Chevignard and Fouque and Schrottenloher (CFS) solved this problem~\cite{chevignard2024reducing}.
They found a way to compute approximate modular exponentiations using only small intermediate values, destroying the anecdotal $n$ qubit arithmetic bottleneck.
However, their method is incompatible with ``qubit recycling''~\cite{Parker2000,Mosca1999}, reviving an old bottleneck on the number of \emph{input} qubits.
Shor's original algorithm used $2n$ input qubits~\cite{shor1994}, but Eker{\aa} and H{\aa}stad proved in 2017 that $(0.5 + \epsilon) n$ input qubits was sufficient for RSA integers~\cite{ekeraa2017quantum}.
So, by combining May et al's result with Eker{\aa} et al's result, the CFS algorithm can factor $n$ bit RSA integers using only $(0.5 + \epsilon) n$ logical qubits~\cite{chevignard2024reducing}.

A notable downside of \cite{chevignard2024reducing} was its gate count.
In \cite{gidney2021factor}, it was estimated that a 2048 bit RSA integer could be factored using 3 billion Toffoli gates and a bit more than $3n$ logical qubits.
Whereas \cite{chevignard2024reducing} uses 2 trillion Toffoli gates and a bit more than $0.5n$ logical qubits.
So \cite{chevignard2024reducing} is paying roughly 1000x more Toffolis for a 6x reduction in space.
This is a strikingly inefficient spacetime trade-off.
However, as I'll show in this paper, the trade-off can be made orders of magnitude more forgiving.
In addition to optimizing the Toffoli count of the algorithm, I'll provide a physical cost estimate showing it should be possible to factor 2048 bit RSA integers in less than a week using less than a million physical qubits (under the assumptions mentioned in the abstract).

The paper is structured as follows.
In \sec{algorithm}, I describe a streamlined version of the CFS algorithm.
In \sec{estimates}, I estimate its cost.
I first estimate Toffoli counts and logical qubit counts, and then convert these into physical qubit cost estimates accounting for the overhead of fault tolerance.
Finally, in \sec{conclusion}, I summarize my results.
The paper also includes \app{mock-ups}, which shows more detailed mock-ups of the algorithm and its physical implementation.

\section{Methods}
\label{sec:algorithm}

\begin{table}[ht]
    \label{tab:notation}
    \include{assets/man/tbl-notation}
\end{table}

In this section, I present a variation of the CFS algorithm.
The underlying ideas are the same, but many of the details are different.
For example, I use fewer intermediate values and I extract the most significant bits of the result rather than the least significant bits.

Beware that, as in \cite{gidney2021factor}, for simplicity, I will describe everything in terms of period finding against $f(e) = g^e \bmod N$ (as in Shor's original algorithm) despite actually intending to use Eker{\aa}-H{\aa}stad-style period finding~\cite{ekeraa2017quantum}.
Ultimately everything decomposes into a series of quantum controlled multiplications, which fundamentally is what is actually being optimized, so what I describe will trivially translate.

\subsection{Approximate Residue Arithmetic}
\label{sec:approx-residue}

Consider an integer $L$ equal to the product of many $\ell$-bit primes from a set $P$:

\begin{equation}
L = \prod_{p \in P} p
\end{equation}

\begin{equation}
    \forall p \in P : (\len p = \ell) \land \text{IsPrime(p)}
\end{equation}

A simple way to do arithmetic on a value modulo $L$ is to store the value as a 2s complement integer in a $\len L$ bit register, and when operating on this register add or subtract multiples of $L$ as appropriate to keep it in the range $[0, L)$.
This requires $\Omega(\log L)$ bits of storage.

Residue arithmetic performs addition and multiplication modulo $L$ by separately performing the operations modulo each of the primes from $P$.
The final result can then be recovered using the Chinese remainder theorem.
By the prime number theorem, $P$ can be chosen so that the bit length of primes in $P$ is $\ell = \Theta(\log \log L)$.
Therefore, using residue arithmetic introduces the possibility of arithmetic being exponentially more space efficient:

\begin{equation}
    \label{eq:define-P}
    |P| = \Theta\left( \frac{\log L}{\log \log L} \right)
\end{equation}

\begin{equation}
    \label{eq:define-ell}
    \ell = \Theta(\log \log L)
\end{equation}

In the context of Shor's algorithm, the key operation we want to compute is a modular exponentiation $V = g^{e} \bmod N$.
Here $g$ is a randomly chosen classical constant in $[2, N)$, $e$ is a uniformly superposed $m = O(\log N)$ qubit integer, and $N$ is the number to factor.
The computation of $V$ can be decomposed into a series of multiplications controlled by the qubits of $e$:

\begin{equation}
    V = g^{e} \bmod N
    = \left(\rangeprod{k}{m} M_k^{e_k}\right) \bmod N
\end{equation}

\begin{equation}
    \label{eq:define-m}
    m = O(\log N)
\end{equation}

In the above equation, $e_k$ is the qubit at little-endian offset $k$ within the register $e$ and $M_k$ is the classically precomputed constant

\begin{equation}
    M_k = g^{(2^{k})} \bmod N
\end{equation}

Because the factors of $N$ aren't known, and because in practice those factors would be large, it's not viable to perform residue arithmetic modulo $N$.
However, performing the arithmetic modulo any other modulus $L \neq N$ creates an issue.
If the accumulating product exceeds $L$, then the register will wrap around.
This shifts its value by a multiple of $L$ and, since $L \bmod N \neq 0$, this offsets the tracked value relative to the true result mod $N$.
The following multiplications would then amplify this error out of control, ruining the computation.
I fix this in a simple way, the same way Chevignard et al fix it, by picking $L$ to be larger than the largest possible product.
This guarantees the wraparound issue never occurs:

\begin{equation}
    \label{eq:bound-L}
    L \geq N^{m}
\end{equation}

With this promise about the size of $L$, we can rewrite the computation of $V$ to use residue arithmetic.
Instead of computing the product directly, we can compute it using a dot product between each residue $r_j$ (the product modulo a prime from $L$) and its contribution factor $u_j$ (the smallest positive integer satisfying $u_j \bmod p_i = \delta_{i,j}$):

\begin{equation}
\label{eq:defrk}
    r_j = \left(\rangeprod{k}{m} M_k^{e_k}\right) \bmod p_j
\end{equation}

\begin{equation}
    u_j = (L / p_j) \cdot \text{MultiplicativeInverse}_{p_j}(L / p_j)
\end{equation}

\begin{equation}
\begin{aligned}
    V &= \left(\rangeprod{k}{m} M_k^{e_k}\right) \bmod L \bmod N
    \\
    &= \left(\rangesum{j}{|P|} r_j u_j \right) \bmod L \bmod N
\end{aligned}
\end{equation}

To simplify later steps, I decompose $r_j$ into its $\ell$ bits $r_{j,k}$, and write $V$ in those terms:

\begin{equation}
    r_j = \rangesum{k}{\ell} r_{j,k} \ll k
\end{equation}

\begin{equation}
\label{eq:comp_v}
    V = \left(\rangesum{j}{|P|} \rangesum{k}{\ell} r_{j,k} \Big[u_j \ll k\Big] \right) \bmod L \bmod N
\end{equation}

The computation of $V$ is now a sum modulo $L$ then modulo $N$, instead of a product modulo $N$.
Crucially, this fixes the issue where errors early in the process would be amplified by later operations.
As a result, we're in a position to start using approximations.

Our goal now is to use approximations to extract the most significant bits of $V$, without computing the entirety of $V$.
To quantify the error in these approximations, we'll track its ``modular deviation'' $\Delta_N(a - b)$.
The modular deviation is the minimum number of increments or decrements needed to turn $a$ into $b$ modulo $N$, divided by $N$:

\begin{equation}
\Delta_N(a - b) = \frac{\min((a - b) \bmod N, (b - a) \bmod N)}{N}
\end{equation}

To approximate additions modulo $N$, we'll truncate all values down to $f$ bits.
When asked to add an offset $s$ into an accumulator, we'll instead add $(s \bmod N) \gg t$ where $t = \len N - f$.
The accumulator will be truncated to $f$ bits, and will be operated modulo $N \gg t$ instead of modulo $N$.
At any time, the approximate result can be extracted by left shifting the accumulator's value by $t$.

Truncating introduces two sources of error.
First, during additions, carries that would have propagated out of the removed part of the register into the kept part of the register will no longer happen.
Second, the accumulating additions wrap around slightly too quickly due to $(N \gg t) \ll t$ being less than $N$.
For an individual truncated addition, both of these errors offset the approximate result by an amount $O(2^t)$, which is a modular deviation of $O(2^{-f})$.
So the modular deviation is exponentially small in the number of kept bits, and will accumulate linearly with the number of operations, meaning a series of $A$ truncated additions has a modular deviation of at most $O(A \cdot 2^{-f})$:

\begin{equation}
    t = (\len N) - f
\end{equation}

\begin{equation}
\label{eq:deviated-sum}
\begin{aligned}
    S &= \left(\rangesum{k}{A} s_k \right) \bmod N
    \\
    \widetilde{S} &= \left(\rangesum{k}{A} (s_k \bmod N) \gg t\right) \bmod (N \gg t)
    \\
    \Delta_N(S - (\tilde{S} \ll t)) &\leq O(A \cdot 2^{-f})
\end{aligned}
\end{equation}

This kind of approximated accumulation isn't quite directly applicable to the current expression for $V$.
The issue is that $V$ isn't being accumulated modulo $N$, it's being accumulated modulo $L$ and only at the end is the modulo $N$ performed.
If we were to accumulate only modulo $N$, instead of modulo $L$ then modulo $N$, then each time the accumulator would have wrapped modulo $L$ we'd miss an offset of $L \bmod N$.
However, $L$ is the product of the primes in our residue system, and we get to choose these primes, so we can pull a trick.
We can optimize $L$ such that $L \bmod N$ has negligible modular deviation.

Numerically, it seems to be the case that picking random sets of small primes results in values of $L \bmod N$ uniformly distributed over the range $[1, N)$.
In cases I've tested, I'm consistently able to find an $L$ with deviation below $2^{-f}$ with high probability by randomly sampling $O(2^f)$ sets of small primes.
I conjecture this is true in general (see \hyperref[assumption1]{Assumption 1}).
I'll show later in \eq{f_size} that the required value of $f$ grows logarithmically with the problem size $\len N$.
Assuming the conjecture and the promise that $f$ will grow like $O(\log \log N)$, an $L$ with sufficiently small deviation can be found in polynomial time:

\begin{equation}
    \label{eq:requireddeviation}
    \Delta_N(L) < 2^{-f} \;\;\;\;\text{(by brute force random search for a satisfying $P$)}
\end{equation}

\begin{mdframed}
    \textbf{Assumption 1 (random search for small modular deviations is efficient)}: A set $P$ of small primes satisfying $(\prod P) \geq N^m$ and $\Delta_N(\prod P) < 2^{-f}$ can be found in $O(2^f \cdot \text{poly}(m \cdot \len N))$ expected time by randomly varying the primes included in $P$.
    \label{assumption1}
\end{mdframed}

As an example, here's a set of 25000 primes I found in ten seconds using a 128 core machine.
Each prime is 22 bits long, and the set achieves a modular deviation below $2^{-32}$ versus the RSA2048 challenge number~\cite{wikirsa2048challengenumber}:

\begin{equation}
    \begin{aligned}
    P_1 &= \text{PrimesBetween}(3814620, 2^{22})
    \\
    P_2 &= \{2097769, 3484783, 3814501, 3814543, 3814561, 3814583, 3814609\}
    \\
    L &= \prod (P_1 \cup P_2)
    \\
    N &= \text{RSA}_{2048}
    \\
    \Delta_{N} (L) &< 2^{-32}
    \end{aligned}
\end{equation}

Given the promise that the modular deviation of the wraparound error is at most $2^{-f}$, we can generalize \eq{deviated-sum} from approximating arithmetic modulo $N$ to approximating arithmetic modulo $L$ then modulo $N$:

\begin{equation}
\label{eq:deviated-sum-2}
\begin{aligned}
    S &= \left(\rangesum{k}{A} s_k \right) \bmod L \bmod N
    \\
    \widetilde{S} &= \left(\rangesum{k}{A} (s_k \bmod L \bmod N) \gg t\right) \bmod (N \gg t)
    \\
    \Delta_N(S - (\tilde{S} \ll t))
    &\leq O(A \cdot 2^{-f} + A \cdot \Delta_N(L))
    \\ &\leq O(A \cdot 2^{-f})
\end{aligned}
\end{equation}

We can now apply \eq{deviated-sum-2} to \eq{comp_v}, producing an expression for $\widetilde{V} \approx V \gg t$:

\begin{equation}
\label{eq:approx-residue}
\begin{aligned}
    \widetilde{V} &= \left(\rangesum{j}{|P|} \rangesum{k}{\ell} r_{j,k} \left(\Big[(u_j \ll k) \bmod L \bmod N \Big] \gg t\right) \right) \bmod (N \gg t)
    \\
    &= \left(\rangesum{j}{|P|} \rangesum{k}{\ell} r_{j,k} C_{j,k} \right) \bmod (N \gg t)
\end{aligned}
\end{equation}

So, by using truncated residue arithmetic, the modular exponentiation has been approximated into an $f$-bit dot product between the bits of the residues and a table of classically precomputable constants $C_{j,k}$:

\begin{equation}
C_{j,k} = \left(\Big[(u_j \ll k) \bmod L \bmod N \Big] \gg t\right) \bmod (N \gg t)
\end{equation}

The computation of $\widetilde{V}$ incurs $O(2^{-f})$ modular deviation per truncated addition, and there are $|P| \cdot \ell$ additions, so the total modular deviation of $\widetilde{V}$ is at most

\begin{equation}
\label{eq:modevbound}
    \Delta_{N}(V - (\widetilde{V} \ll t)) \leq O(|P| \cdot \ell \cdot 2^{-f})
\end{equation}

It will be clear from the next subsection that, in order for period finding to work, it's sufficient to achieve a constant total modular deviation (e.g. a modular deviation of 10\%).
Solving for $f$ in $|P| \cdot \ell \cdot 2^{-f} = O(1)$ after expanding terms using \eq{define-P}, \eq{define-ell}, \eq{define-m}, and \eq{bound-L} gives a bound on the size of $f$:

\begin{equation}
\label{eq:f_size}
    f = O(\log \log N)
\end{equation}

See \fig{example-modexp-code} for working Python code that computes $\widetilde{V} \approx g^e \bmod N$ using this method.

\begin{figure}
    \pythonlisting{7pt}{assets/gen/simple_example_code.py}
    \caption{
    Example Python code that approximates $\widetilde{V} \approx g^e \bmod N$ given a choice of $P$ and $f$.
    Registers that would store quantum values during the quantum factoring are prefixed with ``\texttt{Q\_}''.
    All would-be-quantum registers are of size $O(\log \log N)$, except the input register \texttt{Q\_e}.
    For simplicity, this code omits optimizations (like windowing) and crucial quantum details (like uncomputation).
    See \app{python} for more detailed code.
    }
    \label{fig:example-modexp-code}
\end{figure}

\subsection{Approximate Period Finding}

Shor's algorithm, as usually described, performs period finding against a periodic function $f(x)$.
It relies crucially on the fact that $f(x)$ is \emph{exactly} periodic, with $f(x + P) = f(x)$ for all values of $x$.
In this paper I'm computing an approximation $\widetilde{f} \approx f$, so I can only guarantee \emph{approximate periodicity}:

\begin{equation}
    \forall x, y \in \mathbb{Z} : \Delta_N\left(\widetilde{f}(x + yP) - \widetilde{f}(x)\right) \leq \epsilon.
\end{equation}

Exact period finding will fail if used on a function that only guarantees approximate periodicity, because the error in the approximation can make the function aperiodic.
To fix this, we must avoid measuring the error.
For example, instead of measuring the entire output of $\widetilde{f}(x)$ you could measure a truncated output $\widetilde{f}(x) // (100 N \epsilon)$.
This could work, but causes two problems.
The first problem is that truncating the output means more inputs will be consistent with it.
This will cause the system to collapse to a large superposition of periodic signals, instead of a simple periodic signal.
I'll address this by analyzing the behavior of period finding against superpositions of periodic signals, as was done in \cite{may2022compression}.
The second problem with truncating the output is that the parts of the output that aren't measured would need to be uncomputed, by running the modular exponentiation process backwards.
The modular exponentiation is already the most expensive part of the algorithm, so this would double the execution costs.
I'll address this by using superposition masking.

Superposition masking uses sacrificial superpositions to control the information that can enter into a register or be revealed by measurements~\cite{zalka2006pure,gidney2019approximate,jaques2020masking}.
In the present context, we want to avoid learning the low bits of $\widetilde{f}(x)$.
We can achieve this by measuring $\widetilde{f}(x) + s$ instead of $\widetilde{f}(x)$, where $s$ is a uniform superposition over a contiguous range $[0, \lceil SN \rceil)$.
I'll address the value of $S$ later; for now note that larger values of $S$ will suppress error from the approximation but reduce the amount of information remaining for period finding.

Because $\widetilde{f}(x)$ happens to be computed with a series of additions into an output register, it's possible to compute and measure $\widetilde{f}(x) + s$ without needing to then uncompute $\widetilde{f}(x)$ or $s$.
This is done by initializing the output register to $s$ (instead of 0), then adding the approximation $\widetilde{f}(x)$ into the output register, and then measuring the entire output register.
To understand this process in detail, let's analyze the states that occur as the algorithm progresses.

At the beginning of the algorithm, the output register is storing a uniform superposition (the mask).
There is also the input register: an $m$ qubit register storing its own uniform superposition.
Together these two registers form the initial state $|\psi_0\rangle$:

\begin{equation}
\begin{aligned}
    |\psi_0\rangle
    &=
    |0 : 2^m : 1 \rangle \otimes |0 : \lceil SN \rceil :1\rangle
    \\&=
    \frac{1}{\sqrt{2^m \lceil SN \rceil}}\rangesum{e}{2^m}\rangesum{s}{\lceil SN \rceil} |e\rangle \otimes |s\rangle
\end{aligned}
\end{equation}

The algorithm now adds the approximation $\widetilde{f}(e)$ into the output register, modulo $N$, forming the actual pre-measurement state $|\widetilde{\psi_1}\rangle$:

\begin{equation}
\begin{aligned}
    |\widetilde{\psi_1}\rangle
    &=
    \frac{1}{\sqrt{\lceil SN \rceil 2^m}}\rangesum{e}{2^m}\rangesum{s}{\lceil SN \rceil} |e\rangle \otimes |(s + \widetilde{f}(e)) \bmod N\rangle
\end{aligned}
\end{equation}

If we had used $f$ instead of $\widetilde{f}$, we would have produced the \emph{ideal} pre-measurement state $|\psi_1\rangle$:

\begin{equation}
\begin{aligned}
    |\psi_1\rangle
    &=
    \frac{1}{\sqrt{\lceil SN \rceil 2^m}}\rangesum{e}{2^m}\rangesum{s}{\lceil SN \rceil} |e\rangle \otimes |(s + f(e)) \bmod N\rangle
\end{aligned}
\end{equation}

Consider the infidelity $1 - |\langle \psi_1 |\widetilde{\psi_1}\rangle|^2$.
For each possible computational basis value $e$ of the input register, conditioning $|\widetilde{\psi_1}\rangle$ and $|\psi_1\rangle$ on the input register storing $e$ produces a uniform superposition in the output register covering $\lceil SN \rceil$ contiguous values.
If the maximum modular deviation of $\widetilde{f}$ is $\epsilon$, then the conditioned ranges in $|\widetilde{\psi_1}\rangle$ are offset from the conditioned ranges in $|\psi_1\rangle$ by at most $N\epsilon$.
The width of the ranges is at least $S/\epsilon$ times wider than the offset from the deviation.
Therefore the infidelity between the two conditioned states is a most $\epsilon/S$.
But this is true for \emph{every} condition, and so also bounds the total infidelity between the two states:

\begin{equation}
1 - |\langle \psi_1 |\widetilde{\psi_1}\rangle|^2 \leq \epsilon/S
\label{eq:max-infidelity}
\end{equation}

We can treat this infidelity as contributing a chance of the algorithm failing, and thereby continue the analysis assuming we are working with the ideal state $|\psi_1\rangle$ instead of the actual state $|\widetilde{\psi_1}\rangle$.
For example, if we set $S = 1000 \epsilon$ then $|\widetilde{\psi_1}\rangle$ would have 99.9\% overlap with $|\psi_1\rangle$ and so any success rate analysis we did with $|\psi_1\rangle$ would produce a result within 0.1\% of the true success rate.

The next step in the algorithm, which we'll analyze with $|\psi_1\rangle$ instead of $|\widetilde{\psi_1}\rangle$, is to measure the output register.
This produces some measurement result $V$, and collapses the system into the post-measurement state $|\psi_2\rangle$.
This state is a superposition of all the exponents whose masked output ranges overlapped with $V$:

\begin{equation}
\begin{aligned}
    |\psi_2\rangle
    &\propto
    \rangesum{e}{2^m} |e\rangle \cdot \text{int}\left(0 \leq (V - f(e)) \bmod N < \lceil NS \rceil\right)
\end{aligned}
\end{equation}

Let $P$ be the period of $f$, and let $R$ be the set of exponents less than $P$ that are consistent with the measured value $V$.
We can rewrite $|\psi_2\rangle$ in these terms:

\begin{equation}
\begin{aligned}
    R = \left\{ e \in \mathbb{N} \;\Big|\; e < P \land 0 \leq (V - f(e)) \bmod N < \lceil NS \rceil \right\}
\end{aligned}
\end{equation}

\begin{equation}
\begin{aligned}
    |\psi_2\rangle
    &\approx
    \frac{1}{\sqrt{|R|}}\sum_{e \in R} |e : P : 2^m\rangle
\end{aligned}
\end{equation}

The above equality is approximate because $2^m$ isn't guaranteed to be a multiple of $P$.
Some of the states $|e : P : 2^m\rangle$ have $2^m // P$ non-zero amplitudes while others may have $2^m // P + 1$.
To avoid annoyances like this, I'll continue the analysis while pretending that the signal and quantum Fourier transform are being worked on modulo a multiple of the period ($MP$) rather than modulo a power of 2 ($2^m$).
In reality we must operate modulo $2^m$, since we don't know $P$, but it's well known that the modulo-$2^m$ behavior approximates the modulo-$MP$ behavior~\cite{coppersmith1994qft}.
The error in the approximation decreases exponentially as $m$ increases, because the overlap between a candidate frequency $\widetilde{\theta}$ and a true frequency $\theta$ contains a $2^m$ scaling factor that forces the range of high-overlap candidate frequencies to tighten towards the true frequencies:

\begin{equation}
\begin{aligned}
\\|\theta\rangle &= \frac{1}{\sqrt{2^m}} \sum_{k=0}^{2^m-1} |k\rangle e^{i \theta k}
\\\left| \langle \theta |\widetilde{\theta}\rangle \right|^2
&= \left| \frac{1}{2^m}\sum_{k=0}^{2^m-1} e^{i (\theta - \widetilde{\theta}) k} \right|^2
\approx \text{sinc}^2\left( 2^m (\theta - \widetilde{\theta})/2 \right)
\end{aligned}
\end{equation}

So let's pretend we have a post-measurement state modulo $MP$ instead of modulo $2^m$:

\begin{equation}
\begin{aligned}
    |\psi_2^\prime\rangle
    &=
    \frac{1}{\sqrt{|R|}}\sum_{e \in R} |e : P : MP\rangle
\end{aligned}
\end{equation}

The state $\left|r:MP:P\right\rangle$ is a ``simple periodic signal''.
Shor proved that period finding works on simple periodic signals.
The state $|\psi_2^\prime\rangle$ isn't a simple periodic signal; it's a superposition of such signals.
In \cite{may2022compression} it was shown that period finding works on these states, as long as the contents of $R$ are randomized.
For completeness I'll reproduce a similar argument here, and conjecture that this randomized analysis applies to the actual values of $R$ that appear when factoring.

The modulo-$MP$ quantum Fourier transform of a simple periodic signal $\left|r:MP:P\right\rangle$ is a superposition with non-zero amplitudes at multiples of $MP/P=M$.
The non-zero amplitudes all have equal magnitude, but their phases depend on $r$:

\begin{equation}
    \text{QFT}_{MP} \left|r:MP:P\right\rangle
    = \text{GRAD}_{MP}^r \left|0:MP:M\right\rangle
\end{equation}

Because the QFT of every simple periodic signal only has non-zero amplitudes at multiples of $M$, the QFT of any superposition of simple periodic signals can also only have non-zero amplitudes at multiples of $M$.
No new frequency peaks can appear.
However, there is still the possibility for interference amongst the existing frequency peaks.
If each simple periodic signal $\left|r:MP:P\right\rangle$ is assigned an amplitude $\alpha_r$ then we find:

\begin{equation}
\begin{aligned}
    \text{QFT}_{MP}
    \rangesum{r}{P}
    \alpha_r
    \left|r:MP:P\right\rangle
    &=
    \rangesum{r}{P}
    \alpha_r
    \text{GRAD}_{MP}^r
    \left|0:MP:M\right\rangle
    \\&=
    \rangesum{k}{P}
    \left(
        \frac{1}{\sqrt{P}}
        \rangesum{r}{P}
        \alpha_r
        e^{i 2\pi r k / P}
    \right)
    |kM\rangle
    \\&=
    \rangesum{k}{P}
    \beta_k
    |kM\rangle
\end{aligned}
\end{equation}

where $\beta_k$ is the amplitude of the frequency peak at $|kM\rangle$:

\begin{equation}
    \label{eq:beta_k}
    \beta_k = \frac{1}{\sqrt{P}}
        \rangesum{r}{P}
        \alpha_r
        e^{i 2\pi r k / P}
\end{equation}

In the actual algorithm, the amplitudes of the simple periodic signals are zero for remainders outside a set $R$, and equal for all remainders in $R$:

\begin{equation}
\label{eq:subsetperiodicsuperpose}
    \alpha_r = \frac{\text{int}(r \in R)}{\sqrt{|R|}} 
\end{equation}

Let $w = |R|$ and assume that $R$ was chosen uniformly at random from the $\binom{P}{w}$ possible sets of $w$ remainders modulo $P$.
Our goal is to derive a simple expression for $E(|\beta_k|^2)$, the expected rate of measuring the $k$'th peak, given this assumption.
Start by expanding definitions from \eq{beta_k} and \eq{subsetperiodicsuperpose}, averaged over possible values of $R$:

\begin{equation}
\begin{aligned}
  E(|\beta_k|^2)
  &=
  \frac{1}{\binom{P}{w}}
  \sum_{\substack{R \in \mathcal{P}([0, P)) \\ |R|=w}}
  \left|
    \frac{1}{\sqrt{P}}
    \rangesum{r}{P}
    \frac{\text{int}(r \in R)}{\sqrt{|R|}}
    e^{i k r 2\pi / P}
  \right|^2
  \\&=
  \frac{1}{\binom{P}{w}wP}
  \sum_{\substack{R \in \mathcal{P}([0, P)) \\ |R|=w}}
  \rangesum{r_1}{P}
  \rangesum{r_2}{P}
  \text{int}(r_1 \in R)
  \cdot
  \text{int}(r_2 \in R)
  \cdot
  e^{i k r_1  2\pi / P}
  e^{-i k r_2 2\pi / P}
\end{aligned}
\end{equation}

When $k=0$, all the $e^{ik \dots}$ terms simplify to 1 and the overall expression simplifies to $E(|B_0|^2) = w/P$.
So focus on the case where $k>0$.

The sum over values of $R$ can be pushed rightward, so that it just counts how many values of $R$ are compatible with a choice of $(r_1, r_2)$.
When $r_1 = r_2$, there are $\binom{P-1}{w-1}$ values of $R$ that satisfy $r_1 \in R \land r_2 \in R$.
When $r_1 \neq r_2$, there are $\binom{P-2}{w-2}$ satisfying values.
This replaces the sum over values of $R$ with a choice of multiplier determined by $r_1 \stackrel{?}{=} r_2$:

\begin{equation}
\begin{aligned}
  E(|B_{k>0}|^2)
  &=
  \frac{1}{\binom{P}{w}wP}
  \rangesum{r_1}{P}
  \rangesum{r_2}{P}
  e^{i k r_1  2\pi / P}
  e^{-i k r_2 2\pi / P}
  \left(
  \sum_{\substack{R \in \mathcal{P}([0, P)) \\ |R|=w}}
  \text{int}(r_1 \in R \land r_2 \in R)
  \right)
  \\&=
  \frac{1}{\binom{P}{w}wP}
  \rangesum{r_1}{P}
  \rangesum{r_2}{P}
  e^{ir_1 k 2\pi / P}
  e^{-ir_2 k 2\pi / P}
  \begin{cases}
  r_1=r_2 &\rightarrow \binom{P-1}{w-1}
  \\
  r_1\neq r_2 &\rightarrow \binom{P-2}{w-2}
  \end{cases}
\end{aligned}
\end{equation}

Note that, for non-zero values of $k$, the nested sum in the above equation would evaluate to 0 if the right side multiplier was replaced by a constant:

\begin{equation}
    0 < k < P \implies \rangesum{r_1}{P}\rangesum{r_2}{P} e^{i r_1 k2\pi/P} e^{-i r_2 k2\pi/P} = 0
\end{equation}

This allows the right hand multiplier to be offset by any fixed amount, without changing the total.
I choose to offset by $-\binom{P-2}{w-2}$, zeroing the multiplier for the $r_1 \neq r_2$ case, leaving behind only the $r_1=r_2$ cases.
The expression then simplifies to a fraction independent of $k$:

\begin{equation}
\begin{aligned}
  E(|B_{k>0}|^2)
  &=
  \binom{P}{w}^{-1}
  \frac{1}{wP}
  \rangesum{r}{P}
  e^{ir k 2\pi / P}
  e^{-ir k 2\pi / P}
  \left(\binom{P-1}{w-1}-\binom{P-2}{w-2}\right)
  \\&=
  \binom{P}{w}^{-1}
  \frac{1}{wP}
  P
  \left(\binom{P-1}{w-1}-\binom{P-2}{w-2}\right)
  \\&= \frac{P - w}{P(P-1)}
\end{aligned}
\end{equation}

In other words, when $w$ randomly chosen simple periodic signals are superposed, the system behaves as if $\beta_0$ takes a $w/P$ cut and then $\beta_1, \dots, \beta_{P-1}$ equally split whats left:

\begin{equation}
\begin{aligned}
  E(|\beta_k|^2) =
  \begin{cases}
      k=0 &\rightarrow \frac{w}{P}
      \\
      0 < k < P &\rightarrow \left(1 - \frac{w}{P}\right) \cdot \frac{1}{P-1}
  \end{cases}
\end{aligned}
\end{equation}

Therefore, in the randomized case, period finding against a superposition of $w$ simple periodic signals has a $w/P$ probability of failing due to sampling 0.
In the usual period finding algorithm, the probability of sampling 0 would be $1/P$ instead of $w/P$.
The randomized case otherwise behaves identically to the usual period finding used in Shor's algorithm.

\begin{figure}
    \centering
    \resizebox{\linewidth}{!}{
    \includegraphics{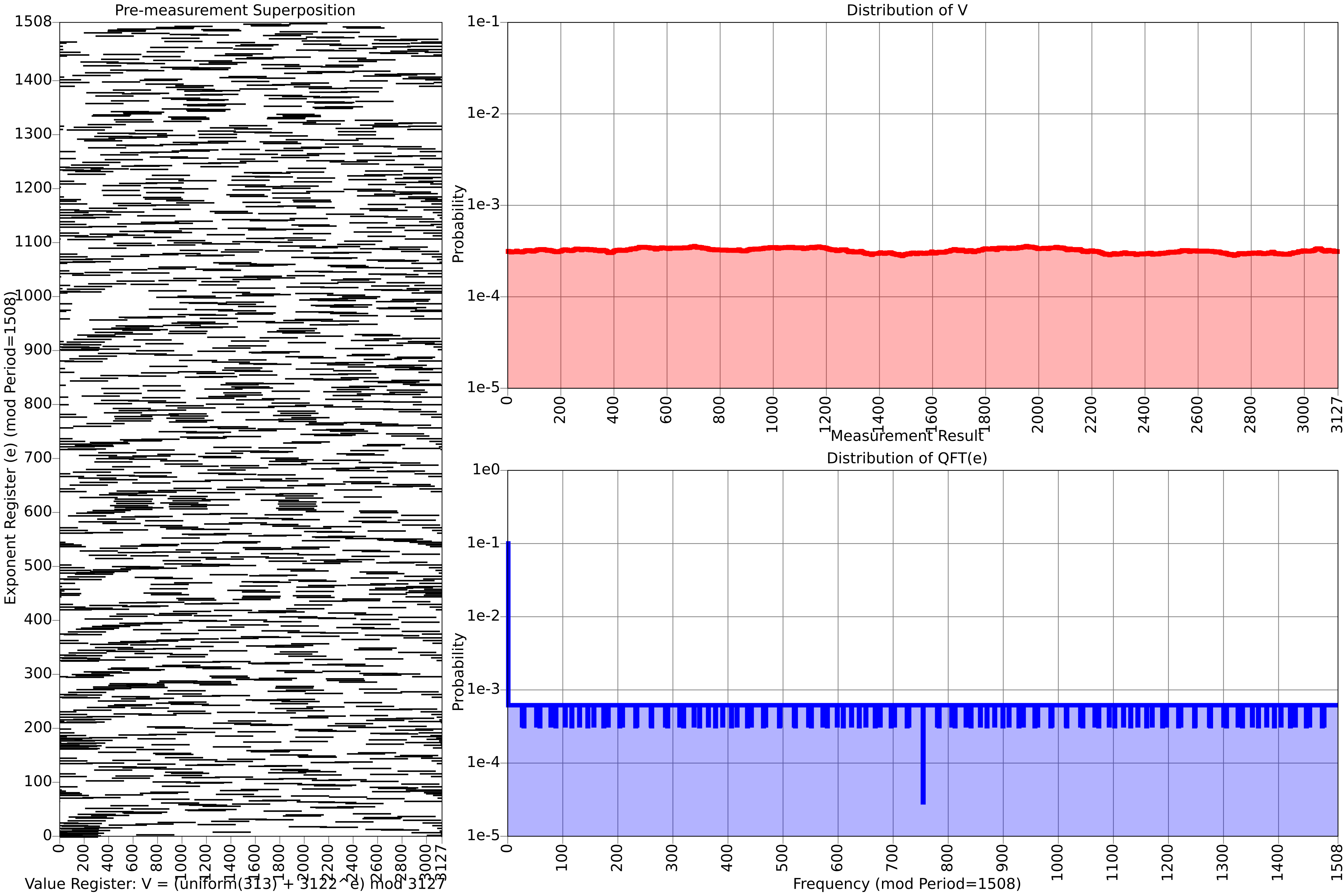}
    }
    \caption{
        A small-scale example of how masking affects period finding.
        Left: a pre-measurement state, with white meaning zero amplitude and black meaning equal non-zero amplitude.
        Top right: probability of measuring different values of the output register.
        Masking has smoothed the distribution to be nearly uniform.
        Bottom right: probability of measuring different frequency peaks of the input register.
        Appears uniform (instead of spiky) because the frequencies are computed modulo the period P.
        Without masking, the bottom right distribution would be exactly uniform.
        With masking, the zero'th frequency is substantially more likely due to constructive interference from the masking.
        Certain frequencies are also less likely.
    }
    \label{fig:compression}
\end{figure}

\begin{figure}
    \centering
    \includegraphics[width=\linewidth]{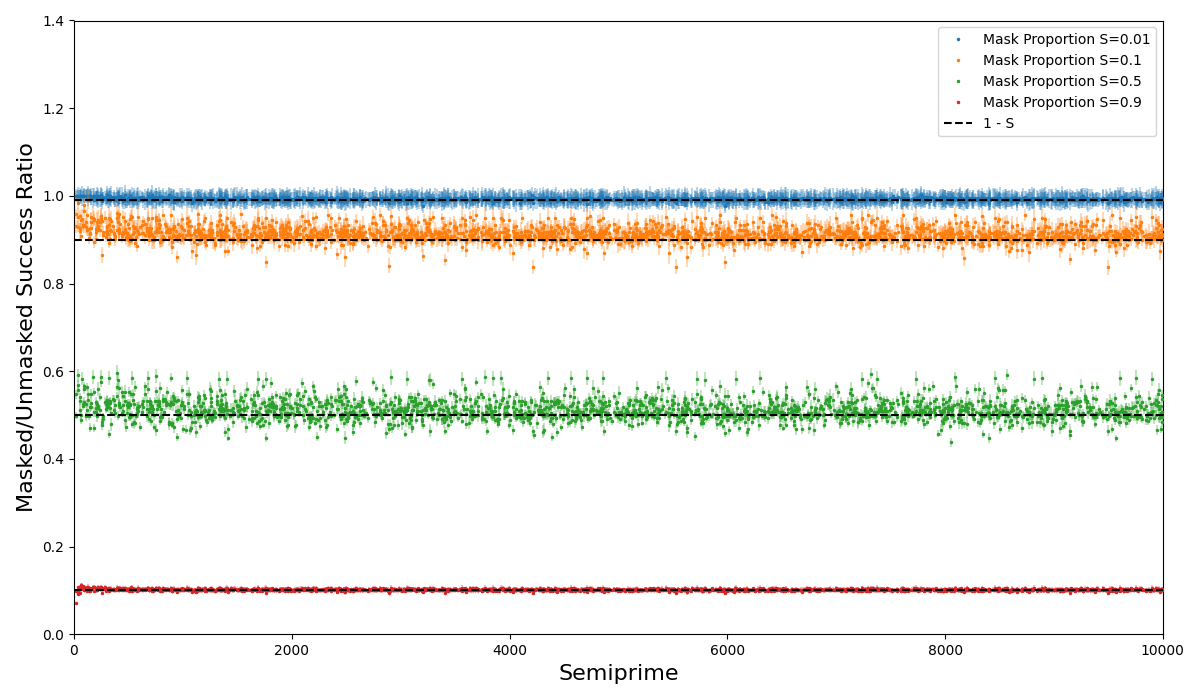}
    \caption{
        Suppression of success rate due to the use of superposition masking, estimated by Monte Carlo sampling.
        Error bars are computed by separately computing likelihoods for the masked samples and the unmasked samples, finding the respective high/low hypothesis probabilities that have a likelihood within 100x of the max likelihood hypothesis given the sampled data, and then making the smallest/largest ratios possible with those high and low values.
    }
    \label{fig:masked-success-rate}
\end{figure}

Of course, in an actual factoring, the set $R$ isn't random.
It's a deterministic function of $N$, $g$, and the sampled measurement.
The success rate won't necessarily be suppressed by the factor $1 - w/P \approx 1-S$ that it would be in the random case.
In \fig{compression}, I show an example case.
You can see in that figure that the frequency spectrum has some dips.
These wouldn't be present in the randomized case, and will skew the success rate.
In \fig{masked-success-rate}, I show success rate numerics for different values of $S$ and $N$.
The first thing I notice when looking at the data is that the success rate appears to skew slightly higher than $1 - S$.
Some instances are slightly worse than $1-S$ but most are slightly better.
The average success rate is higher than I expected.
It also seems that the variation in success rate is small (there's no instance where a masking proportion of $0.1$ produced a success suppression factor below $1 - 0.2$ or where a masking proportion of $0.01$ produced a success suppression factor below $1-0.03$).
When collecting data for the figure, I was hoping the variations wouldn't just be small but would noticeably decrease as $N$ was increased.
No such effect is apparent in the data (keeping mind that the plotted values of $N$ are tiny compared to $2^{2048}$).

Although the numerics show that the real case isn't exactly identical to the randomized case, they appear close enough that, for my purposes in this paper, the real costs can be estimated by using the randomized analysis.
I will estimate costs as if a superposition mask proportion of $S$ incurs a probability $S$ of the shot failing and a probability $1-S$ of the shot behaving as if no masking was present (see \hyperref[assumption2]{Assumption 2}).

\begin{minipage}{0.95\textwidth}
\begin{mdframed}
    \textbf{Assumption 2 (behavior of masked period finding)}:
    When using a superposition mask that covers a proportion $S$ of the output space, the cost of using masking can be roughly estimated by multiplying the expected number of shots by $1/(1-S)$.
    \label{assumption2}
\end{mdframed}
\end{minipage}
\vspace{1em}

Let's now return to choosing the proportional width $S$ of the superposition mask.
Recall from \eq{max-infidelity} that the maximum infidelity of the analysis, caused by using an approximation $\widetilde{f}$ with a modular deviation of at most $\epsilon$, is $\epsilon/S$.
Also recall from \hyperref[assumption2]{Assumption 2} that a mask covering a proportion $S$ of the output space behaves like an $S$ chance of failure.
The sum $S + \epsilon/S$ of these two error mechanisms is an upper bound on the chance $P_{\text{deviant}}$ of a shot failing due to using masking and approximations:

\begin{equation}
    P_{\text{deviant}} \leq S + \epsilon/S
\end{equation}

Minimizing this sum produces a choice for $S$, the proportional size of the mask:

\begin{equation}
    S = \sqrt{\epsilon}
\end{equation}

This implies that an approximation with modular deviation $\epsilon$ causes a failure rate of at most

\begin{equation}
    P_{\text{deviant}} \leq 2\sqrt{\epsilon}
\end{equation}

This could likely be improved, for example by using a Gaussian mask instead of a uniform mask or by analyzing the distribution of deviations rather than focusing on the worst case deviation.
But $2\sqrt{\epsilon}$ suffices for my purposes in this paper, so I leave such optimizations to future work.

\subsection{Eker\aa-H{\aa}stad Period Finding}

Elsewhere in the paper, I describe computations in terms of period finding against the function $f(e) = g^e \pmod{N}$.
This would require an input register with $2n$ qubits~\cite{shor1994}, where $n = \len N$.
As in Chevignard et al's paper, I actually use Eker\aa-H{\aa}stad-style period finding~\cite{ekeraa2017quantum} instead of Shor-style period finding.
Eker\aa-H{\aa}stad-style period finding is specialized to the RSA integer case (it requires that $N$ factors into two similarly-sized primes), but uses $n/2 + n/s$ input qubits instead of $2n$.

Eker\aa-H{\aa}stad-style period finding works as follows.
The algorithm is given a positive integer parameter ($s$).
A classical value $g$ is chosen, uniformly at random, from $\mathbb{Z}_N^\ast$ (the multiplicative group modulo $N$).
A second classical value $h = g^{N - 1} \bmod N$ is derived from $g$ and $N$.
An $n/2 + \lceil n/(2s) \rceil$ qubit register ($x$) is prepared into a uniform superposition.
An $\lceil n/(2s) \rceil$ qubit register ($y$) is similarly prepared into a uniform superposition.
The value $g^x h^y \pmod{N}$ is computed under superposition, and discarded.
Then $x$ and $y$ are measured in the frequency basis, producing a data point $(x^\prime, y^\prime)$.
This is repeated at least $s+1$ times.
Post-processing then recovers, with high probability, $d = \text{dlog}_{g \pmod{N}}(h)$ from the list of recorded points.
With high probability it will be the case that $d = p+q-2$, where $p, q$ are the prime factors, because

\begin{equation}
\begin{aligned}
    g^d  \bmod N &= h \bmod N
    \\&= g^{N-1} \bmod N
    \\&= g^{pq-1} \bmod N
    \\&= g^{(p-1)(q-1) + p + q -2} \bmod N
    \\&= g^{p + q - 2} \bmod N
\end{aligned}
\end{equation}

The factors are then recovered by solving for $p$ in the quadratic equation $p \cdot (d - p + 2) = N$.

A detailed analysis of the post-processing and the expected number of repetitions is available in \cite{ekera2020postprocess} (see \href{https://link.springer.com/article/10.1007/s10623-020-00783-2/tables/1}{table 1 of that paper} in particular).
In this paper, I'll stay in the regime where the expected number of repetitions is $s+1$.
And of course $g^x h^y \pmod{N}$ will be computed approximately, instead of exactly, since it compiles into a series of controlled multiplications the same way $g^e \bmod N$ would have.

Because Eker\aa-H{\aa}stad-style period finding involves combining multiple shots, I should discuss how to handle bad shots.
Bad shots caused by masking are usually easy to detect, because the most common failure should be the quantum computer returning the integer 0 rather than a useful value.
On the other hand, bad shots caused by a logical gate error during the computation are essentially silent.
They manifest as the postprocessing failing despite having collected $s+1$ shots.
When this occurs, as long as errors are bounded to reasonable rates, it's sufficient to take a few extra shots while running the postprocessing on all $\genfrac(){0pt}{2}{t}{s+1}$ possible combinations of shots (where $t$ is the total number of shots).
If $t$ grows too large to reasonably check all the combinations, restart with a different choice of $g$.

Recall from \eq{bound-L} that $L$, the size of the residue system, must be at least $N^m$ (where $m$ is the number of multiplications or equivalently the number of input qubits).
Because increasing $s$ reduces the number of input qubits, it reduces the number of multiplications which then reduces the number of primes in the residue system which then reduces the total amount of work that needs to be done.
This compounded benefit notably improves the spacetime tradeoff of increasing $s$.
For example, using $s=3$ results in a computation that is both smaller \emph{and} faster than using $s=1$ despite the overhead of taking additional shots.

\subsection{Arithmetic Optimizations}
\label{sec:arithmetic-optimizations}

The example code I showed in \fig{example-modexp-code} is simple, but inefficient.
In this subsection, I describe optimizations that reduce its cost by orders of magnitude.

The first optimization is to replace modular multiplications with modular additions, by classically precomputing short discrete logarithms.
This optimization was introduced in \cite{chevignard2024reducing}, but I describe it here for completeness.

Recall that, for each prime $p$ in the residue system, we need to quantum compute a controlled product to get $p$'s residue $V_p = V \bmod p$:

\begin{equation}
    V_p = \prod_{k=0}^{m-1} M_k^{e_k} \pmod{p}
\end{equation}

Because $p$ is small (e.g. 22 bits long), it's feasible to find a multiplicative generator $g_p$ modulo $p$ and to solve for $D_{p,k}$ in

\begin{equation}
    g_p^{D_{p,k}} \bmod p = M_k \bmod p
\end{equation}

Note that this equation doesn't have a solution when $M_k \bmod p = 0$.
In \cite{chevignard2024reducing}, they track the values of $p$ and $M_k$ for which this occurs, and handle them as a special case on the quantum computer.
I instead avoid this corner case by making it a selection criteria of the residue number system that it doesn't occur.
That is to say, I require that

\begin{equation}
    \label{eq:primecondition}
    \forall p \in P, \forall x \in M : p \bmod x \neq 0
\end{equation}

where $P$ is the desired set of primes making up the residue system and $M$ is a given set of all possible multipliers.
This criterion implies the residue system can only be chosen \emph{after} picking the random generator $g$ used by Shor's algorithm, since $M$ depends on $g$.
($M$ also depends on windowing parameters I'll describe later.)
Consequently, I use different residue systems even when targeting the same value of $N$.
Choosing the residue system is a per-shot cost, not a per-factoring cost.

To ensure it's possible to efficiently find a satisfying residue system, primes failing \eq{primecondition} need to be excluded \emph{before} making the random variations mentioned in \hyperref[assumption1]{Assumption 1}.
In the rare event that too many primes are excluded, resulting in a search space so small that a value of $P$ satisfying \eq{requireddeviation} is unlikely to exist, just increment the target prime bit length $\ell$.
This will roughly double the number of candidate primes, and roughly halve the chance of each candidate prime failing \eq{primecondition}, and so should quickly guarantee a solution exists if repeated.

Every value $D_{p,k} = \text{dlog}(g_p, M_k, p)$ can be precomputed before starting the quantum computation.
As a result, the controlled product computation is replaced by a controlled sum computation:

\begin{equation}
    S_p = \sum_{k=0}^{m-1} D_{p,k} \cdot e_k
\end{equation}

You can then compute $V_p$ from $S_p$ using a modular exponentiation.
To further reduce costs, use the fact that the order of $g_p$ is known to be $p-1$ and compress $S_p$ into $S_p \bmod (p-1)$ before exponentiating:

\begin{equation}
    V_p = g_p^{S_p \bmod (p - 1)} \bmod p
\end{equation}

Note that this exponentiation performs multiplications, which is the operation we were trying to avoid.
However, crucially, the number of multiplications is now $\len p$ (i.e. dozens) instead of $O(\log N)$ (i.e. thousands).

The second optimization that I apply is windowing~\cite{kutin2006containswindowing,vanmeter2005fastmodexp,gidney2019windowedarithmetic}.
When computing $S_{p_j}$ from $e$, iterate over $e$ in chunks of $w_1$ qubits at a time instead of 1 qubit at a time.
The $w_1$ qubits are used as the address of a lookup into a classical table of $2^{w_1}$ values, each being the discrete logarithm of the combined product that would have been multiplied into the total if the qubits took on a specific value.
Performing the lookup has a cost of $2^{w_1}$, but reduces the number of iterations of the inner loop computing $S_{p_j}$ from $m$ to $\lceil m/w_1 \rceil$.

Similar to increasing the Eker\aa-H{\aa}stad parameter $s$, increasing $w_1$ gives compounded benefits.
A larger value of $w_1$ batches more multiplications together, which reduces the effective number of multiplications, which by \eq{bound-L} reduces the required size of the residue system.

I also use windowing when computing $V_{p_j}$ from $S_{p_j}$.
The method is essentially identical to what's shown in \cite{gidney2019windowedarithmetic}, except that two multiplications are saved by initializing the accumulator using a lookup.
This was independently suggested in \cite{luongo2025arithmetic}.

I also use windowing when adding approximate offsets into the final total, but only because lookups were needed in that loop anyways.
It has a negligible impact on the overall cost of the algorithm.
Lookups could also be used when reducing $S_{p_j}$ modulo $p_j$, with similarly negligible impact.

The third optimization I apply is to merge the computation of $S_{p_j}$ with the uncomputation of $S_{p_{j-1}}$.
In the transition from iteration $j-1$ to iteration $j$, I could uncompute $S_{p_{j-1}}$ by a series of controlled subtractions $\rangesum{k}{m} -D_{p_{j-1},k-1} \cdot e_k$ and then compute $S_j$ by a series of controlled additions $\rangesum{k}{m} D_{p_j,k} \cdot e_k$.
But, because these use the same controls $e_k$, these two things can be merged.
$S_{p_{j-1}}$ can be mutated into $S_{p_{j}}$ by performing controlled additions of precomputed per-exponent-qubit differences $D_{p_j,k} - D_{p_{j-1},k}$:

\begin{equation}
D^\prime_{j,k} = D_{p_j,k} - D_{p_{j-1},k}
\end{equation}

\begin{equation}
S_{p_j} = S_{p_{j-1}} + \rangesum{k}{m} D^\prime_{p_j,k} \cdot e_k
\end{equation}

This is a simple optimization, significant only because computing each value of $S_{p_j}$ is a substantial portion of the cost of the algorithm.

The fourth important optimization that I apply is deferring and merging phase corrections.
In many places in the algorithm, measurement based uncomputations produce phasing tasks.
These phasing tasks don't need to be performed immediately, and often the phasing tasks produced in one place can be merged with phasing tasks produced in another place.
I do this so many times throughout the implementation that it would be tedious to list them all.
But, for example, consider that a lookup performed by unary iteration~\cite{babbush2018} has AND gate uncomputations that correspond to CZ gates.
The later uncomputation of the output of the lookup will produce substantially more complex phase corrections~\cite{berry2019qubitization}.
The CZ corrections produced by the computation are a subset of the phase corrections that can occur during the uncomputation, and the CZ corrections can be deferred until the uncomputation.
So the CZ corrections can be merged into the uncomputation.
From a cost perspective, this deletes the CZ corrections.

A fifth optimization I apply is to flip each modular addition \texttt{X += T[b] (mod p)} into a subtraction \texttt{X -= T2[b] (mod p)} (where \texttt{T2[b] = p - T[b]} is a trivially modified lookup table).
The benefit of doing this is that an addition exceeding \texttt{p} requires an additional comparison to detect, but a subtraction underflowing 0 is easy to detect by concatenating an extra qubit \texttt{Q} to the top of the \texttt{X} register.
\texttt{Q} will flip to 1 iff the subtraction underflows.
So, after the subtraction, the modular addition can be completed by performing \texttt{X += [0, p][Q]} and then uncomputing \texttt{Q} using measurement based uncomputation.
The uncomputation requires a phase correction 50\% of the time, where the phase correction is based on a comparison.
This modular adder costs $2.5n \pm O(1)$ Toffolis gates in expectation (compared to $3.5n \pm O(1)$ in \cite{berry2024rapidprep}).
If the modular addition is later uncomputed, its cost can be further reduced to $2n \pm O(1)$ by deferring and merging the phase correction.

See \app{python} for reference Python code implementing the optimized arithmetic.

\section{Results}
\label{sec:estimates}

\subsection{Logical Costs}

Essentially all work performed by the algorithm is addition, lookup, and ``phaseup'' operations:

\begin{itemize}
\item An $n$ qubit addition is an operation that acts on an $n$ qubit offset register ($a$) and an $n$ qubit target register ($b$).
It performs the classical transition $(a, b) \rightarrow (a, (b + a) \bmod 2^n)$ and has a Toffoli cost of $n - 1$ as well as an ancilla cost of $n-1$~\cite{gidney2018addition}.
Note that, for the purposes of cost estimation, subtractions and comparisons are counted as additions because their circuits are nearly identical.
\item A lookup is an operation, parameterized by a classical table of values $T$, that acts on an $n$ qubit address register ($a$) and a quantum output register.
It performs the classical transition $(a, 0) \rightarrow (a, T_a)$ and has a Toffoli cost of $2^n - n - 1$ as well as an ancilla cost of $n-1$~\cite{babbush2018}.
\item A ``phaseup'' is an operation, parameterized by a classical table of values $T$, that acts on an $n$ qubit target register ($a$).
It negates the amplitudes of states that have a non-zero entry in the table, and has a Toffoli cost of $\sqrt{2^n} \pm O(n)$~\cite{berry2019qubitization}.
\end{itemize}

To estimate total cost, I first estimate the numbers and sizes of these three operations by counting occurrences in the reference code in \app{python}.
\tbl{subroutine-tallies} shows symbolic tallies of these quantities.
I also count qubit allocations and deallocations, to list symbolic qubit tallies in \tbl{subroutine-qubit-tallies}.
The symbolic tallies can be turned into numeric estimates by choosing parameter values (see \tbl{symbols}) and substituting.

To pick parameters, I ran a grid scan.
The Eker\aa-H{\aa}stad parameter $s$ was ranged from 2 to 14.
The prime bit length $\ell$ used by the residue arithmetic system was ranged from 18 to 25.
The window size $w_1$ used by loop1 was ranged from 2 to 8.
The window size $w_3=w_{3a}=w_{3b}$ used by loop3 and unloop3 was ranged from 2 to 6.
The window size $w_4$ used by loop4 was ranged from 2 to 8.
The length $f$ of the truncated accumulator was ranged from 24 to 59.
Infeasible combinations were discarded (e.g. if the prime bit length is too small, it won't be possible to find a set of primes $\prod P \geq N^{m/w_1}$).

The best performing parameter combinations are shown as Pareto frontiers in \fig{pareto-curves}.
I also chose to highlight, in \tbl{logical-costs}, parameters that minimized $q^3 t$ where $q$ is the qubit count and $t$ is the Toffoli count.
Cubing $q$ is just an arbitrary way of favoring space savings more than time savings.

\begin{table}[ht]
    \centering
\resizebox{\linewidth}{!}{
\begin{tabular}{|c|c|l|l|}
     \hline Symbol & Equivalent to & Asymptotic & Description
     \\\hline\hline $N$ & - & $\theta(2^n)$ & Number to factor.
     \\\hline $n$ & $\len N$ & $\theta(n)$ & Bit size of number to factor.
     \\\hline $\ell$ & - & $\theta(\log n)$ & Bit size of primes in residue system.
     \\\hline $s$ & - & $\theta(1)$ & Eker\aa-H{\aa}stad parameter.
     \\\hline $m$ & $\lceil n/2 + n/s \rceil$ & $\theta(n)$ & Number of input qubits.
     \\\hline$f$ & - & $\theta(\log n)$ & Size of output accumulator.
     \\\hline$w_1$ & $\log_2(\ell + \len m) \pm O(1)$ & $\theta(\log \log n)$ & Window length used by loop 1.
     \\\hline$w_3$ & $\log_2(\ell)/2 \pm O(1)$ & $\theta(\log \log n)$ & Window length used by loop 3 and unloop 3.
     \\\hline$w_4$ & $\log_2(\ell) \pm O(1)$ & $\theta(\log \log n)$ & Window length used by loop 4.
     \\\hline$W_1$ & $\lceil m / w_1 \rceil$ & $\theta(n/\log \log n)$ & Number of windows iterated by loop 1.
     \\\hline$W_3$ & $\lceil \ell / w_3 \rceil$ & $\theta(\log n / \log \log n)$ & Number of windows iterated by loop 3.
     \\\hline$W_4$ & $\lceil \ell / w_4 \rceil$ & $\theta(\log n / \log \log n)$ & Number of windows iterated by loop 4.
     \\\hline $|P|$ & $\approx nm/(\ell \cdot w_1)$ & $\theta(n^2 / (\log n \log \log n))$ & Number of primes in residue system.
     \\\hline
\end{tabular}
}
\caption{Descriptions of variables used in cost estimates.}
\label{tbl:symbols}
\end{table}

\begin{table}[ht]
    \centering
\resizebox{\linewidth}{!}{
\begin{tabular}{|l|l|c|c|c|c|c|l|}
\hline
Subroutine & Iterations & Register Size & Address Size & Additions & Lookups & Phaseups & Toffolis
\\\hline\hline
loop1 & $(|P| + 1) \cdot W_1$ & $\ell + \len m$ & $w_1$ & 1 & 1 &0& $\widetilde{\Theta}(n^3)$
\\\hline
loop2 & $|P| \cdot \len m$ & $\ell + \len m$ & - & 2 & 0 & 0& $\widetilde{\Theta}(n^2)$
\\\hline
      loop3 (startup)& $|P|$ &$\ell$& $2 w_3$  & 0& 1 &0&  $\widetilde{\Theta}(n^2)$
\\
loop3 (body) & $|P| \cdot (W_3 - 2) \cdot W_3$ & $\ell$ & $w_3$ & 2 & 1 &0 &  $\widetilde{\Theta}(n^2)$
\\\hline
loop4 & $|P| \cdot W_4$ & $f$ & $w_4$ & 1.5 & 2.5 & 1 &  $\widetilde{\Theta}(n^2)$
\\\hline
unloop3 (body) & $|P| \cdot (W_3 - 2) \cdot 2 \cdot W_3$ & $\ell$ & $w_3$ & 2.5 & 1.5 & 1 &  $\widetilde{\Theta}(n^2)$
\\
unloop3 (cleanup) & $|P|$ &$\ell$& $2 w_3$  & 0  & 0& 1 &  $\widetilde{\Theta}(n^2)$
\\\hline
unloop2 & $|P| \cdot \len m$ & $\ell + \len m$ & - & 2 &  0 &0&  $\widetilde{\Theta}(n^2)$
\\\hline
\end{tabular}
}
\caption{
    Symbolic tallies of additions, lookups, and phaseups used by the subroutines of the algorithm.
    The variables being used are defined in \tbl{symbols}.
}
\label{tbl:subroutine-tallies}
\end{table}

\begin{table}[ht]
    \centering
\resizebox{\linewidth}{!}{
\begin{tabular}{|l|l|l|l|}
\hline
Subroutine & Added Qubits & Temporary Qubits & Total Qubits
\\\hline\hline
Algorithm Startup & $m + f$ & $0$ & $m + f$
\\\hline
Enter Outer Loop & $\ell + \len m$ & $0$ & $m + f + \ell + \len m$
\\\hline
\;\;\;\;loop1 & $0$ & $2(\ell + \len m)$ & $m + f + 3 \ell + 3~\len m$
\\\hline
\;\;\;\;loop2 & $0$ & $\ell + \len m$ & $m + f + 2 \ell + 2~\len m$
\\\hline
\;\;\;\;loop3 & $\ell$ & $2\ell$ & $m + f + 4 \ell + \len m$
\\\hline
\;\;\;\;loop4 & $0$ & $2f$ & $m + 3 f + 2 \ell + \len m$
\\\hline
\;\;\;\;unloop3 & $-\ell$ & $2\ell$ & $m + f + 4 \ell + \len m$
\\\hline
\;\;\;\;unloop2 & $-0$ & $\ell + \len m$ & $m + f + 2 \ell + 2~\len m$
\\\hline
Exit Outer Loop & $-\ell - \len m$ & $2(\ell + \len m)$ & $m + f + 3\ell + 3~\len m$
\\\hline
Measure Approximate Result & $-f$ & $0$ & $m + f$
\\\hline
Frequency Measurement & $-m$ & $O(\log(1/\epsilon))$ & $m + O(\log(1/\epsilon))$
\\\hline
\end{tabular}
}
\caption{
    Symbolic tallies of qubits used by parts of the algorithm.
    The variables being used are defined in \tbl{symbols}.
}
\label{tbl:subroutine-qubit-tallies}
\end{table}

\begin{table}[ht]
    \centering
    \resizebox{\linewidth}{!}{
    \input{assets/gen/logical-cost-table}
    }
    \caption{
        Highlighted parameter choices, and resulting logical cost estimates, for factoring RSA integers of various sizes.
        $P_{\text{deviant}}$ is the shot failure rate due to using an approximate modular exponentiation with masking.
        E(shots) is the expected number of shots, equal to $(s+1)/(1 - P_{\text{deviant}})/0.99$ (the $/0.99$ accounts for the chance of postprocessing failure~\cite{ekera2020postprocess}).
        The Toffolis column is expected Toffolis per factoring (not per shot).
    }
    \label{tbl:logical-costs}
\end{table}

\FloatBarrier
\subsection{Physical Costs}

The imagined physical layout of the algorithm is as follows.
There will be three regions: compute, hot storage, and cold storage.
The hot storage region will store qubits ``normally'', as distance $d$ surface code patches using $2(d+1)^2$ physical qubits per logical qubit.
The cold storage region will store qubits more densely, by using yoked surface codes~\cite{gidney2024yoked}.
The compute region will have room for performing lattice surgery and use magic state cultivation~\cite{gidney2024cultivation} followed by 8T-to-CCZ distillation~\cite{jones2013toffoli} to power Toffoli gates.

\begin{figure}
    \centering
    \resizebox{\linewidth}{!}{
    \includegraphics{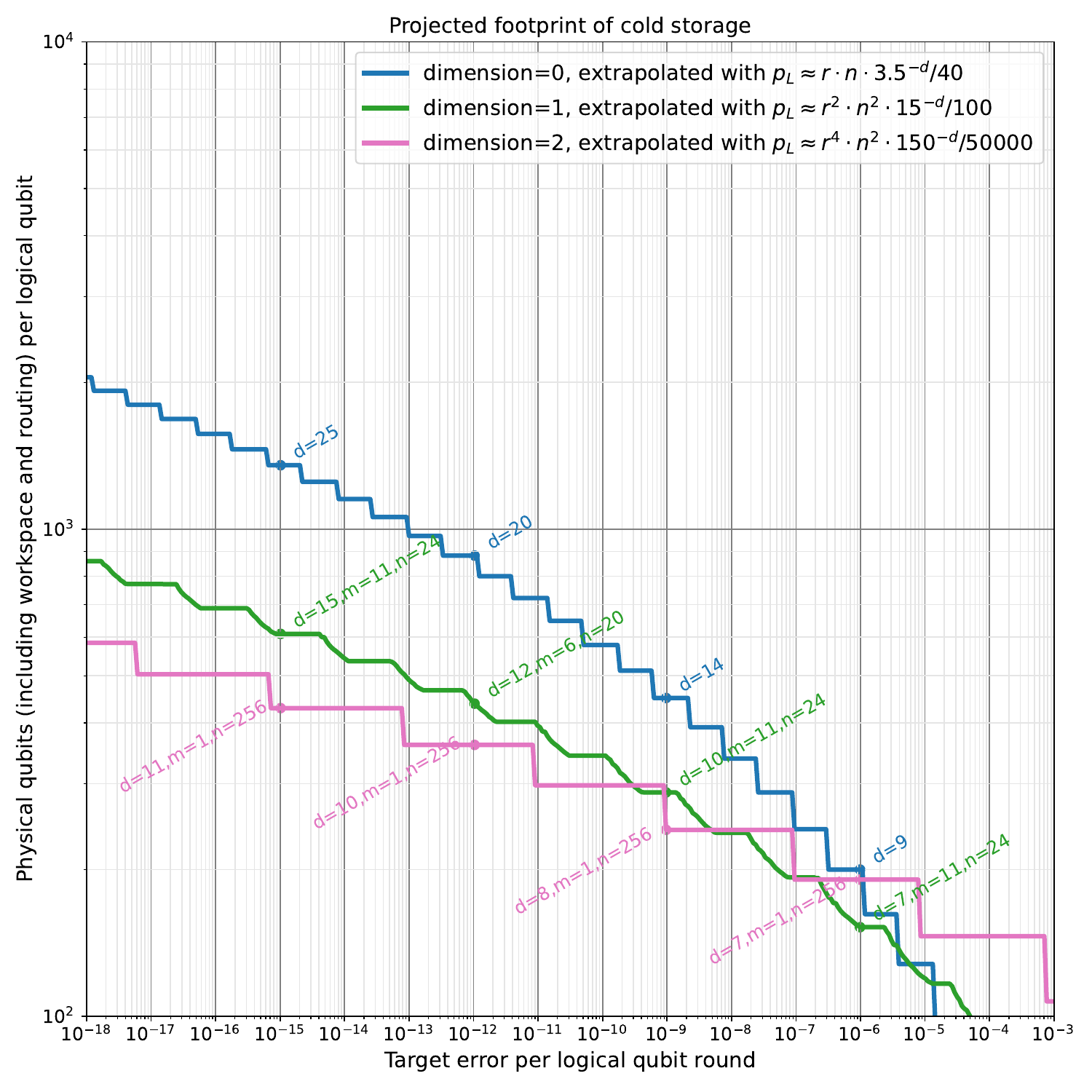}
    }
    \caption{
        Projections of the cost of storage of normal surface codes and surface codes ``yoked by'' (concatenated below) 1D and 2D parity check codes, under uniform depolarizing noise.
        The blue line is storage using normal surface codes (``hot storage'' in this paper).
        The pink line is storage using surface codes yoked by a two dimensional parity check code (``cold storage'' in this paper).
        The $d,m,n$ labels indicate the surface code patch diameter $(d)$, the number of used surface code patches per yoked code instance $(n)$, and the number yoked code instances sharing lattice surgery workspace $(m)$.
        This is a variant of \href{https://static-content.springer.com/esm/art\%3A10.1038\%2Fs41467-025-59714-1/MediaObjects/41467_2025_59714_MOESM1_ESM.pdf\#page=2}{supplementary figure 2 of} \cite{gidney2024yoked}, with fit parameters rounded to one-and-a-half sig figs instead of one sig fig.
        I'm using this variation because I was worried that the projected costs with one sig fig were too optimistic, due to rounding up the fit constants.
        Specifically: in this figure the dimension=0 suppression gain parameter is 3.5 instead of 4, and the dimension=2 suppression gain parameter is 150 instead of 200, making the curves more pessimistic.
    }
    \label{fig:yoked-costs}
\end{figure}

I will show later in this subsection that, when factoring a 2048 bit integer, each shot will take roughly 12 hours and involve fewer than 1600 logical qubits (including idle hot patches).
Given the assumed surface code cycle time of 1 microsecond, this implies $1600 \cdot 12 \cdot 60 \cdot 60 \cdot 10^6 \approx 6.9 \cdot 10^{13}$ logical qubit rounds of runtime to protect.
Choosing a target logical error rate of $10^{-15}$ per logical qubit round will thus result in a no-logical-error shot rate of $(1 - 10^{-15})^{6.9 \cdot 10^{13}} \approx 93.3\%$.

Referring to \fig{yoked-costs}, note that a distance of 25 is sufficient for normal surface code patches to reach a per-patch per-round logical error rate of $10^{-15}$.
So my hot patches will use $2 \cdot (25+1)^2 = 1352$ physical qubits per logical qubit.
For the cold storage, again referring to \fig{yoked-costs}, yoking with a 2D parity check code reaches a logical error rate of $10^{-15}$ when using 430 physical qubits per logical qubit.
So cold logical qubits will be roughly triple the density of hot logical qubits.

The $n=2048$ row of \tbl{logical-costs} specifies $s=8$, $\ell=21$, $w_1=6$, $w_3=3$, $w_4=5$, $f=33$, and $m=1280$.
Plugging these values into \tbl{subroutine-qubit-tallies}, the maximum logical qubit count occurs during loop4 when there are $m+3f+2\ell+\len m=1409$ logical qubits active.
The $m=1280$ input logical qubits are in cold storage, and so cover $1280 \cdot 430 = 550400$ physical qubits.
The remaining $3f+2\ell+\len m=131$ logical qubits are in hot storage, and so cover $131 \cdot 1352 = 177112$ physical qubits.
Finally, the compute region will use a 7x18 region of hot patches ($170352$ physical qubits).
This is enough for six magic state factories, each covering a $3 \times 4$ area of hot patches, as well as three columns of workspace.
The factories (from \href{https://arxiv.org/pdf/2409.17595#page=28}{figure 24 of} \cite{gidney2024cultivation}) are notably smaller than the $15 \times 8$ factories used in \cite{gidney2021factor}, due to replacing the first stage of distillation with magic state cultivation.
The total number of logical qubits, including idle hot patches, is $1280 + 131 + 7 \cdot 18 = 1537 < 1600$, as promised when picking the code distances.
The total number of physical qubits is $897864$.

For slack, I report the physical qubit count as one million instead of 900 thousand.
Specifically, I'm a bit worried that the cold storage might get slightly worse during loop1 of the algorithm.
During this loop, the yoke measurements would have to contend with Z-splitting copies~\cite{de2017zxlattice} of cold logical qubits to stream to hot storage to be used as the addresses of lookups.
I'm confident 100K physical qubits is sufficient slack to ensure the streaming can be interleaved with the required yoke checks.
I leave detailed analysis and simulations of cold storage, under various workloads, as future work.

\begin{figure}[ht]
    \centering
    \resizebox{\linewidth}{!}{
    \includegraphics{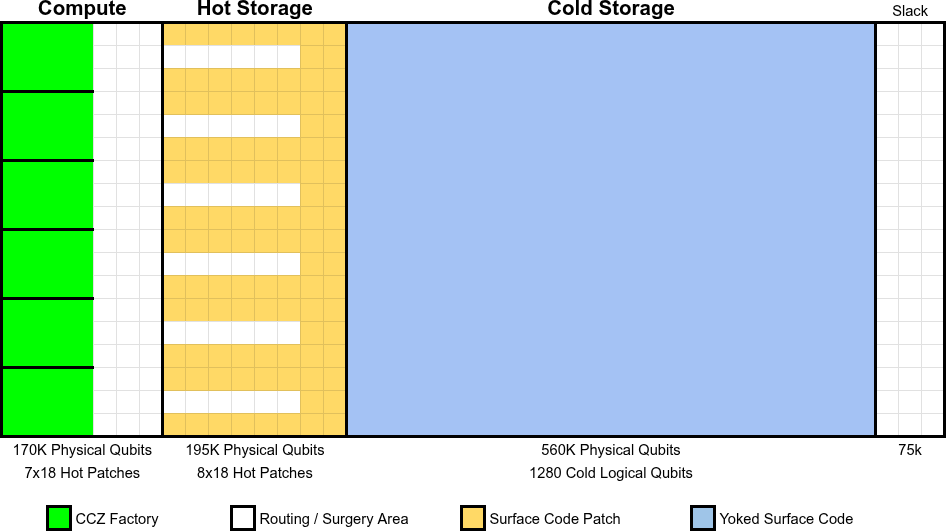}
    }
    \caption{
        Mock-up of a physical layout for $n=2048,s=8$.
        On the left: a narrow ``compute'' region containing six magic state factories and some workspace for lattice surgery.
        On the right: a large ``cold storage'' region for high-density storage of idle qubits using yoked surface codes~\cite{gidney2024yoked}
        In the middle: ``hot storage'' for active qubits that need to rapidly interact with the compute region.
        Some slack space is included in case of unforeseen inefficiencies.
    }
    \label{fig:space-overview}
\end{figure}

A mock-up of the spatial layout (after some rounding) is shown in \fig{space-overview}.

Let's now turn to runtime.
According to \href{https://arxiv.org/pdf/2409.17595#page=2}{figure 1 of} \cite{gidney2024cultivation}, cultivating a T state with a logical error rate of $10^{-7}$ uses $30000$ physical qubit~$\cdot$~rounds of spacetime volume.
Each 8T-to-CCZ factory covers $3 \cdot 4 \cdot 26^2 \cdot 2$ qubits, and needs 8 T states, suggesting an average cultivation time of $14.7$ rounds.
The factory itself (shown in \href{https://arxiv.org/pdf/2409.17595#page=28}{figure 24 of} \cite{gidney2024cultivation}) has 6 layers of lattice surgery.
It uses temporally encoded lattice surgery~\cite{chamberland2022tels} so each layer should be able to execute in $2/3d$ rounds rather than $d$ rounds.
In total this amounts to 114.7 rounds per CCZ state, which I round up to 150 rounds for slack.
Since the magic state cultivation targets a logical error rate of $10^{-7}$ and 8T-to-CCZ distillation has an error suppression of $28p^2$~\cite{jones2013toffoli}, the distilled CCZ states will have a logical error rate of $28 \cdot (10^{-7})^2 < 10^{-12}$.
This is low enough (compared to the Toffoli count of 6.5 billion and other error rates in the paper) that we can ignore it.
Note that, because there are six CCZ factories, the lattice surgery period ($25$ microseconds because $d=25$) happens to be equal to the CCZ state period ($25$ microseconds because $150/6=25$).
These two numbers (combined with \tbl{subroutine-tallies}) allow me to bound how long additions, lookups, and phaseups take.

The largest additions that occur are the $f=33$ qubit additions into the output register during loop4.
These additions require $32$ CCZ states.
It will take at least $32 \cdot 25 = 800$ microseconds to generate and consume these states.
After these states are consumed, the adder has finished computing its output but it still needs to uncompute one of its inputs.
This is strictly simpler, not requiring any CCZ states, so the whole adder will take at most $800 \cdot 2 = 1600$ microseconds.
I round this up to 2 milliseconds to leave slack for things like rearranging qubits at the beginning and end of the operation.
See \app{addition-operation} for a more detailed mock-up of the addition operation.

The largest lookups that occur are the $w_1=6$ address qubit lookups during loop1.
These lookups need at least $2^6-6-1=57$ CCZ states, and at least $63$ layers of lattice surgery~\cite{babbush2018}.
This will take at least $63 \cdot 25 = 1575$ microseconds.
I again round this up to 2 milliseconds for slack.
See \app{lookup-operation} for a more detailed mock-up of the lookup operation.

The largest phaseup operations also use 6 qubit addresses.
This requires 8 CCZ states (instead of the 64 used by a 6 qubit lookup), and the amount of lattice surgery required is similarly more than twice as small as what's needed for a lookup.
So we can safely conclude phaseup operations will take less than half as long as lookups; at most 1 millisecond.
See \app{phaseup-operation} for a more detailed mock-up of the phaseup operation.

With operation durations in hand, we can plug parameter values into  \tbl{subroutine-tallies} and then add up the iterations times the operations times the durations.
Using 2 milliseconds per addition, 2 milliseconds per lookup, and 1 millisecond per phaseup produces an estimate of 12.07 hours per shot (as promised when picking the code distance).
The $n=2048$ row of \tbl{logical-costs} indicates the expected number of shots is $9.2$, giving an expected total time of $12.07 \cdot 9.1 / 24 = 4.63$ days.
However, this doesn't yet account for shots lost due to logical errors.
Recall from earlier that shots have a 93.3\% chance of not having a logical error.
Dividing $4.63$ days by $93.3\%$ gives the actual time estimate per factoring: $4.96$ days.
Which I round up to a week for slack.

So, summarizing, I estimate that factoring a 2048 bit RSA integer requires less than a million noisy qubits and will finish in less than a week.
This estimate assumes a quantum computer with a surface code cycle time of 1 microsecond, a control system reaction time of 10 microseconds, a square grid of qubits with nearest neighbor connectivity, and a uniform depolarizing noise model with a noise strength of 1 error per 1000 gates.
See \fig{historical-progression-physical} for a comparison of this estimate to previous cost estimates.

\FloatBarrier
\section{Conclusion}
\label{sec:conclusion}

In this paper, I reduced the expected number of qubits needed to break RSA2048 from 20 million to 1 million.
I did this by combining and streamlining results from \cite{chevignard2024reducing}, 
\cite{gidney2024yoked}, and \cite{gidney2024cultivation}.
My hope is that this provides a sign post for the current state of the art in quantum factoring, and informs how quickly quantum-safe cryptosystems should be deployed.

Without changing the physical assumptions made by this paper, I see no way reduce the qubit count by another order of magnitude.
I cannot plausibly claim that a 2048 bit RSA integer could be factored with a hundred thousand noisy qubits.
But there's a saying in cryptography: ``attacks always get better''~\cite{schneierattacksgetbetter}.
Over the past decade, that has held true for quantum factoring.
Looking forward, I agree with the initial public draft of the NIST internal report on the transition to post-quantum
cryptography standards~\cite{nist2024}: vulnerable systems should be deprecated after 2030 and disallowed after 2035.
Not because I expect sufficiently large quantum computers to exist by 2030, but because I prefer security to not be contingent on progress being slow.

\section{Acknowledgments}

I thank Greg Kahanamoku-Meyer, Thiago Bergamaschi, Dave Bacon, Cody Jones, James Manyika, and Matt McEwen for helpful discussions.
I thank Michael Newman for discussions and for revisiting \cite{gidney2024yoked} to produce the variant of one its figures shown in \fig{yoked-costs}.
I thank Adam Zalcman for discussions and for contributing code to speed up finding prime sets with low modular deviation.
I thank Martin Eker{\aa} for helpful discussions, as well as for helpful corrections, and I look forward to Martin extending the results to other cryptographic cases and becoming a co-author in a future version of this paper.
I thank Hartmut Neven, and the Google Quantum AI team as a whole, for creating an environment where this work was possible.

\printbibliography

\clearpage
\appendix

\section{Detailed Mock-ups}
\label{app:mock-ups}

\subsection{Reference Python Implementation of Algorithm}
\label{app:python}

In this section, I provide tested reference Python code for the optimized algorithm.
Understanding the code requires knowing its conventions.
In the example code, all quantum values are stored as variables of type \texttt{quint}.
These variables are prefixed with \texttt{Q\_} and highlighted red.
A quint is a superposed unsigned integer with a known length in qubits (accessed via python's \texttt{len} function).
Quints can be sliced as if they were Python lists, which returns a view of a subsection of the quint.
For example, if \texttt{Q\_example} is a quint then \texttt{Q\_example[1:5]} is a view of its second, third, fourth, and fifth qubits (the view's integer value is equal to \texttt{Q\_example // 2 \% 16}).
Expressions like \texttt{(Q\_a << len(Q\_b)) | Q\_b} are also just constructing views, by concatenating qubits, not performing actual work on the quantum computer.

Quints are created by allocation calls like \texttt{qpu.alloc\_quint} or \texttt{qpu.alloc\_phase\_gradient}, and cleared by ``del'' calls like \texttt{qpu.del\_by\_equal\_to}.
Measurement based uncomputations begin with a call to \texttt{qpu.mx\_rz} or \texttt{qpu.del\_measure\_x}.
For each qubit in the quint, these methods measure the qubit in the X basis then either clear the qubit to $|0\rangle$ or deallocate it.
The measurements determine if parts of the superposition where the corresponding qubit was ON had their amplitude negated by phase kickback, and are bit packed into an \texttt{int} returned by the method.

The simulation code is expected to uncompute all registers and kickback phases before finishing, which is verified by calling \texttt{qpu.verify\_clean\_finish()} (not shown in the example code).
Under the hood, the simulator works by tracking the value and phase of a few randomly sampled classical trajectories.
This isn't sufficient to verify interference effects, or to verify that information wasn't incorrectly revealed (e.g. it can't verify that masking was done correctly), but it fuzzes that the classical output is correct and that phase kickback from measurement based uncomputation is being fixed.

The key quantum operations that appear in the example code are additions, subtractions, lookups, and phase flip comparisons.
An addition looks like \texttt{Q\_a += b}, which offsets \texttt{Q\_a} by \texttt{b} (working modulo \texttt{2**len(Q\_a)}).
A subtraction looks like \texttt{Q\_a -= b}.
A phase flip comparison looks like \texttt{qpu.z(Q\_a < b)}, where the \texttt{qpu.z} is shorthand for ``phase flip the parts of the superposition where''.
Additions, subtractions, and comparisons all have circuits that are minor variations on an addition circuit and so, in cost estimates, they are all treated as additions.

Lookups appear as parts of other expressions.
For example, \texttt{Q\_a += T[Q\_b]} implies that a lookup of the table \texttt{T} addressed by \texttt{Q\_b} will be performed in order to produce the offset value that will be added into \texttt{Q\_a}.
The lookup is uncomputed by the end of the addition using measurement based uncomputation, producing phase flip corrections to do.
These phase flip corrections go into the ``vent'' of the lookup, which will be specified earlier in the code with a line like \texttt{T = T.venting\_into(V)}.
Later in the code, a call like \texttt{qpu.z(V[Q\_b])} will appear, which is the phaseup used to resolve the accumulated phase corrections from the lookups.
Alternatively, a call like \texttt{qpu.push\_uncompute\_info(V)} may appear.
This call will be matched by a \texttt{qpu.pop\_uncompute\_info()} call in a later subroutine, which will handle the phase corrections.
This is done because often an address is used for multiple lookups, for example during a computation subroutine and again during an uncomputation subroutine, and the phase corrections of such lookups can be merged.

A special kind of lookup that appears is GHZ lookups like \texttt{Q\_a += Q\_b.ghz\_lookup(k)}.
This is a lookup with a single address qubit and a lookup table of the form \texttt{[0, k]} for some integer $k$.
These lookups can be implemented very efficiently in lattice surgery, by Z-splitting~\cite{horsman2012latticesurgery} the single address qubit into each non-zero value of $k$.
At the end of the lookup the qubits can be Z-merged by measuring all but one of them in the X basis, and applying a corrective Z gate to the remaining qubit if an odd number of the measurements returned \texttt{True}.
So a GHZ lookup is more akin to a layer of lattice surgery than to a full blown table lookup, and correspondingly I don't count GHZ lookups as lookups in cost estimates.

Much of the code refers to an instance of \texttt{ExecutionConfig}, which includes information about the exponentiation that is being performed as well as resources like the precomputed tables of numbers used by the various subroutines.
Readers who want to see the details of this class or run the code should refer to \href{https://doi.org/10.5281/zenodo.15347487}{the Zenodo upload}~\cite{gidneyzenodo2025factor}.

\pythonlisting{7pt}{assets/gen/detailed_example_code.py}

\subsection{Addition Operation}
\label{app:addition-operation}

Given two inputs $a$ and $b$, their sum $s=a+b$ satisfies this equation at each bit position $k$:

\begin{equation}
\label{eq:adder-identity}
\begin{aligned}
    s_{k+1} &= \left((a_k \oplus s_k) \cdot (b_k \oplus s_k)\right) \oplus (a_k \oplus b_k \oplus s_k \oplus a_{k+1} \oplus b_{k+1})
\end{aligned}
\end{equation}

Note that the identity is built out of three Z parity products: $P_1 = Z_{a_k} Z_{s_k}$, $P_2 = Z_{b_k} Z_{s_k}$, and $P_3 = Z_{a_k}Z_{b_k}Z_{s_k}Z_{a_{k+1}}Z_{b_{k+1}}$.
This makes the identity interesting, when using lattice surgery, because Z parity products are relatively cheap to access~\cite{fowler2018latticesurgery,litinski2018}.

\begin{figure}[ht]
    \centering
    \resizebox{\linewidth}{!}{
    \includegraphics{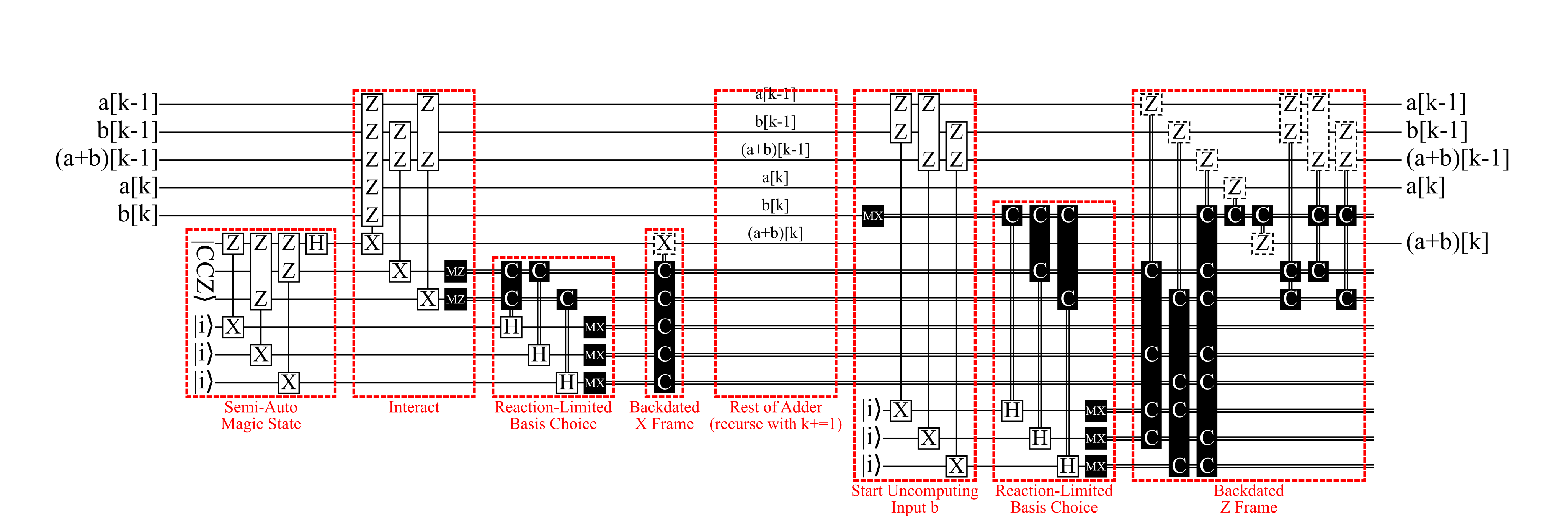}
    }
    \caption{
        A quantum circuit diagram of an adder including Clifford corrections.
        Boxes containing X/Z characters are multi-qubit Pauli product operators.
        Classical parities are shown using ``C'' operators.
        Boxes linked by vertical lines form controlled operations, meaning they only act on their common -1 eigenspaces.
        \href{
            https://algassert.com/quirk\#circuit=\%7B\%22cols\%22\%3A\%5B\%5B1,1,1,1,1,\%22\%E2\%80\%A2\%22,\%22\%E2\%80\%A2\%22,\%22Z\%22\%5D,\%5B\%22~lbl0\%22,\%22~lbl1\%22,\%22~lbl2\%22,\%22~lbl3\%22,\%22~lbl4\%22\%5D,\%5B1,1,1,1,1,\%22Z\%22,1,1,\%22xpar\%22\%5D,\%5B1,1,1,1,1,\%22zpar\%22,1,\%22zpar\%22,1,\%22X\%22\%5D,\%5B1,1,1,1,1,\%22zpar\%22,\%22zpar\%22,1,1,1,\%22X\%22\%5D,\%5B1,1,1,1,1,\%22H\%22\%5D,\%5B\%22zpar\%22,\%22zpar\%22,\%22zpar\%22,\%22zpar\%22,\%22zpar\%22,\%22X\%22\%5D,\%5B1,\%22zpar\%22,\%22zpar\%22,1,1,1,\%22X\%22\%5D,\%5B\%22zpar\%22,1,\%22zpar\%22,1,1,1,1,\%22X\%22\%5D,\%5B1,1,1,1,1,1,\%22Measure\%22,\%22Measure\%22\%5D,\%5B1,1,1,1,1,1,\%22zpar\%22,\%22zpar\%22,\%22H\%22\%5D,\%5B1,1,1,1,1,1,\%22zpar\%22,1,1,\%22H\%22\%5D,\%5B1,1,1,1,1,1,1,\%22zpar\%22,1,1,\%22H\%22\%5D,\%5B1,1,1,1,1,1,1,1,\%22H\%22,\%22H\%22,\%22H\%22\%5D,\%5B1,1,1,1,1,1,1,1,\%22Measure\%22,\%22Measure\%22,\%22Measure\%22\%5D,\%5B1,1,1,1,1,\%22X\%22,\%22zpar\%22,\%22zpar\%22,\%22zpar\%22,\%22zpar\%22,\%22zpar\%22\%5D,\%5B1,1,1,1,\%22H\%22\%5D,\%5B1,1,1,1,\%22Measure\%22\%5D,\%5B\%22zpar\%22,\%22zpar\%22,1,1,1,1,1,1,1,1,1,\%22X\%22\%5D,\%5B\%22zpar\%22,1,\%22zpar\%22,1,1,1,1,1,1,1,1,1,\%22X\%22\%5D,\%5B1,\%22zpar\%22,\%22zpar\%22,1,1,1,1,1,1,1,1,1,1,\%22X\%22\%5D,\%5B1,1,1,1,\%22zpar\%22,1,1,1,1,1,1,\%22H\%22\%5D,\%5B1,1,1,1,\%22zpar\%22,1,\%22zpar\%22,1,1,1,1,1,\%22H\%22\%5D,\%5B1,1,1,1,\%22zpar\%22,1,1,\%22zpar\%22,1,1,1,1,1,\%22H\%22\%5D,\%5B1,1,1,1,1,1,1,1,1,1,1,\%22H\%22,\%22H\%22,\%22H\%22\%5D,\%5B1,1,1,1,1,1,1,1,1,1,1,\%22Measure\%22,\%22Measure\%22,\%22Measure\%22\%5D,\%5B\%22Z\%22,1,1,1,1,1,\%22zpar\%22,1,1,\%22zpar\%22,1,\%22zpar\%22,\%22zpar\%22\%5D,\%5B1,\%22Z\%22,1,1,1,1,1,\%22zpar\%22,1,1,\%22zpar\%22,\%22zpar\%22,1,\%22zpar\%22\%5D,\%5B1,1,\%22Z\%22,1,\%22zpar\%22,1,\%22zpar\%22,\%22zpar\%22,1,\%22zpar\%22,\%22zpar\%22,1,\%22zpar\%22,\%22zpar\%22\%5D,\%5B1,1,1,\%22Z\%22,\%22zpar\%22\%5D,\%5B1,1,1,1,\%22zpar\%22,\%22Z\%22\%5D,\%5B\%22Z\%22,\%22Z\%22,1,1,1,1,\%22\%E2\%80\%A2\%22,\%22\%E2\%80\%A2\%22\%5D,\%5B\%22Z\%22,1,\%22Z\%22,1,\%22\%E2\%80\%A2\%22,1,\%22\%E2\%80\%A2\%22\%5D,\%5B1,\%22Z\%22,\%22Z\%22,1,\%22\%E2\%80\%A2\%22,1,1,\%22\%E2\%80\%A2\%22\%5D,\%5B\%22~lbl0\%22,\%22~lbl1\%22,\%22~lbl2\%22,\%22~lbl3\%22,1,\%22~lbl5\%22\%5D\%5D,\%22init\%22\%3A\%5B0,0,0,0,0,\%22+\%22,\%22+\%22,\%22+\%22,\%22i\%22,\%22i\%22,\%22i\%22,\%22i\%22,\%22i\%22,\%22i\%22\%5D,\%22gates\%22\%3A\%5B\%7B\%22id\%22\%3A\%22~lbl0\%22,\%22name\%22\%3A\%22a\%5Bk\%2D1\%5D\%22,\%22matrix\%22\%3A\%22\%7B\%7B1,0\%7D,\%7B0,1\%7D\%7D\%22\%7D,\%7B\%22id\%22\%3A\%22~lbl1\%22,\%22name\%22\%3A\%22b\%5Bk\%2D1\%5D\%22,\%22matrix\%22\%3A\%22\%7B\%7B1,0\%7D,\%7B0,1\%7D\%7D\%22\%7D,\%7B\%22id\%22\%3A\%22~lbl2\%22,\%22name\%22\%3A\%22\%28a+b\%29\%5Bk\%2D1\%5D\%22,\%22matrix\%22\%3A\%22\%7B\%7B1,0\%7D,\%7B0,1\%7D\%7D\%22\%7D,\%7B\%22id\%22\%3A\%22~lbl3\%22,\%22name\%22\%3A\%22a\%5Bk\%5D\%22,\%22matrix\%22\%3A\%22\%7B\%7B1,0\%7D,\%7B0,1\%7D\%7D\%22\%7D,\%7B\%22id\%22\%3A\%22~lbl4\%22,\%22name\%22\%3A\%22b\%5Bk\%5D\%22,\%22matrix\%22\%3A\%22\%7B\%7B1,0\%7D,\%7B0,1\%7D\%7D\%22\%7D,\%7B\%22id\%22\%3A\%22~lbl5\%22,\%22name\%22\%3A\%22\%28a+b\%29\%5Bk\%5D\%22,\%22matrix\%22\%3A\%22\%7B\%7B1,0\%7D,\%7B0,1\%7D\%7D\%22\%7D\%5D\%7D
        }{Click here to open this circuit in Quirk.}
    }
    \label{fig:adder-circuit}
\end{figure}

A circuit diagram based on \eq{adder-identity} is shown in \fig{adder-circuit}.
The first half of the diagram is an out-of-place adder.
It computes another qubit of $s$, using qubits from $a$ and $b$.
The second half uncomputes a qubit of $b$, using the fact that $b=s-a$.
Note that the diagram inlines details of a teleportation used to perform a Toffoli gate.
Normally, teleported Toffolis are corrected using CX gates and CZ gates~\cite{jones2013toffoli,gidney2019autoccz}.
The circuit instead applies corrections based on teleporting S gates into multi-qubit Pauli products.
Also, the circuit defers corrections that only affect phases until uncomputing $b = a - s$.
This allows those corrections to be merged into similar corrections during the uncomputation, balancing out the circuit's usage of delayed choice routing qubits.
It uses 3 routing qubits during the computation and also 3 during the uncomputation, instead of 6 during the computation and 2 during the uncomputation~\cite{gidney2019autoccz}.

\begin{figure}[ht]
    \centering
    \resizebox{\linewidth}{!}{
    \includegraphics{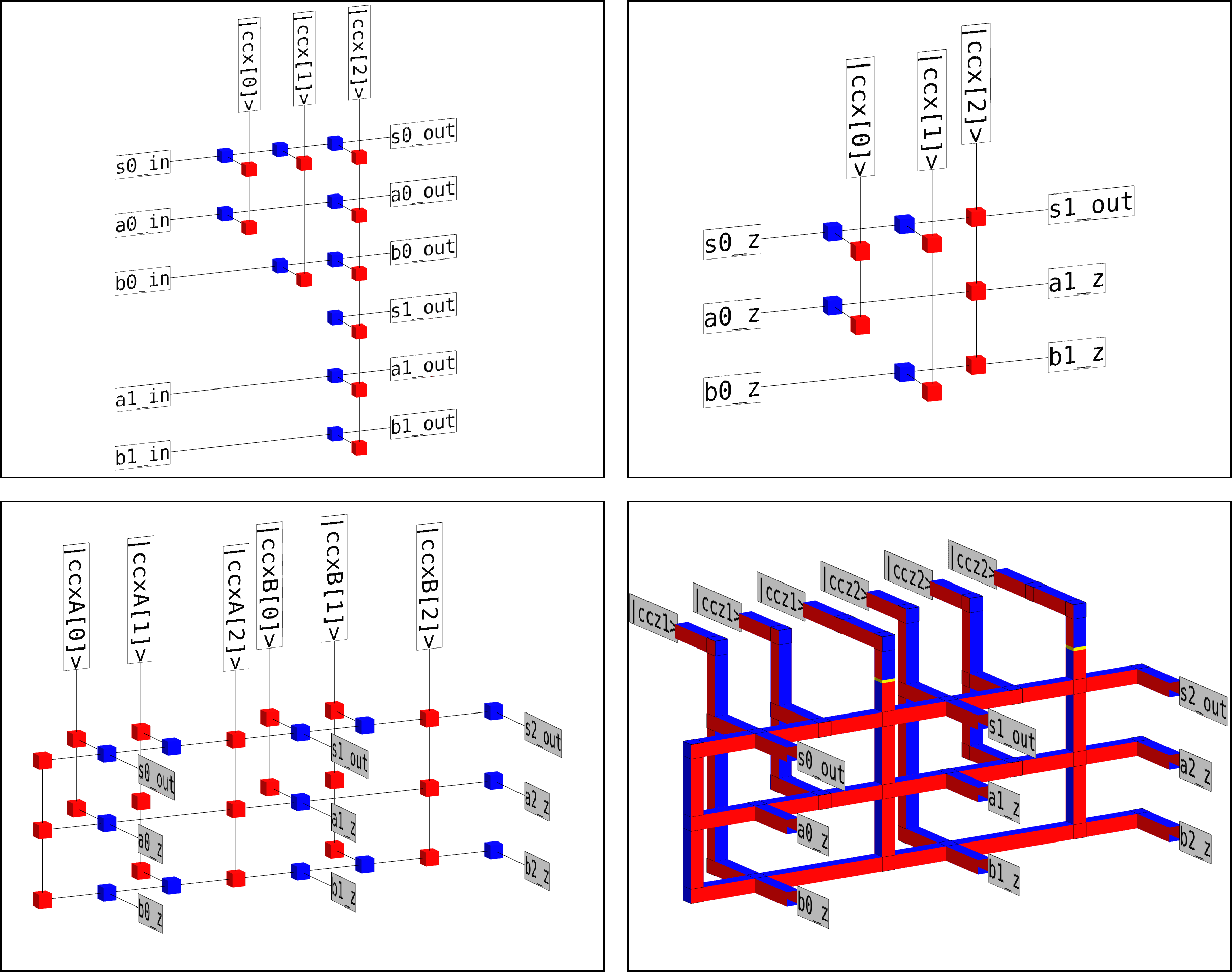}
    }
    \caption{
        Adder building block diagrams.
        Top left is a ZX graph implementing \eq{adder-identity}.
        Top right is the graph after merging input/output ports into Z ports.
        Bottom left is a complete 3-qubit adder built by attaching together multiple instances of the graph.
        Bottom right is an equivalent lattice surgery diagram.
    }
    \label{fig:adder-layout}
\end{figure}

\fig{adder-layout} shows how \eq{adder-identity} can be implemented as a ZX graph and then optimized into an efficient lattice surgery layout.
(The diagram doesn't include the delayed choice correction operations shown in \fig{adder-circuit}.
They're attached to the CCZ state before it enters the picture.)
An important detail here is that the optimized graph has a different ``calling convention'' from the initial graph.
Instead of the qubits of $a$ and $b$ being passed into the adder via input ports, and then returned via output ports, they are attached to the adder via a ``Z port''.
To pass a qubit into a Z port, Z-split the qubit into two qubits~\cite{de2017zxlattice} then Z-merge one of them into the Z port.
The remaining qubit stays behind, becoming the output qubit without moving.
This calling convention is beneficial because it halves the number of ports.

The physical arrangement of the addition is as follows.
The two registers to add are arranged on opposite sides of short hallways leading to the compute region.
Conveniently, these hallways have a pitch of 3 surface code patches which matches the pitch of the adder building block and also the pitch of the magic state factories.
The qubits within the hallways are ordered so that the least significant qubits are further away from the compute region.
As the adder block executes, the hallway is used to access the next two input qubits and then the adder block produces a qubit of the sum, which is stored in the hallway between the two inputs used to produce it.
The hallway fills in gradually as the sum is computed.

\begin{figure}[ht]
    \centering
    \resizebox{\linewidth}{!}{
    \includegraphics{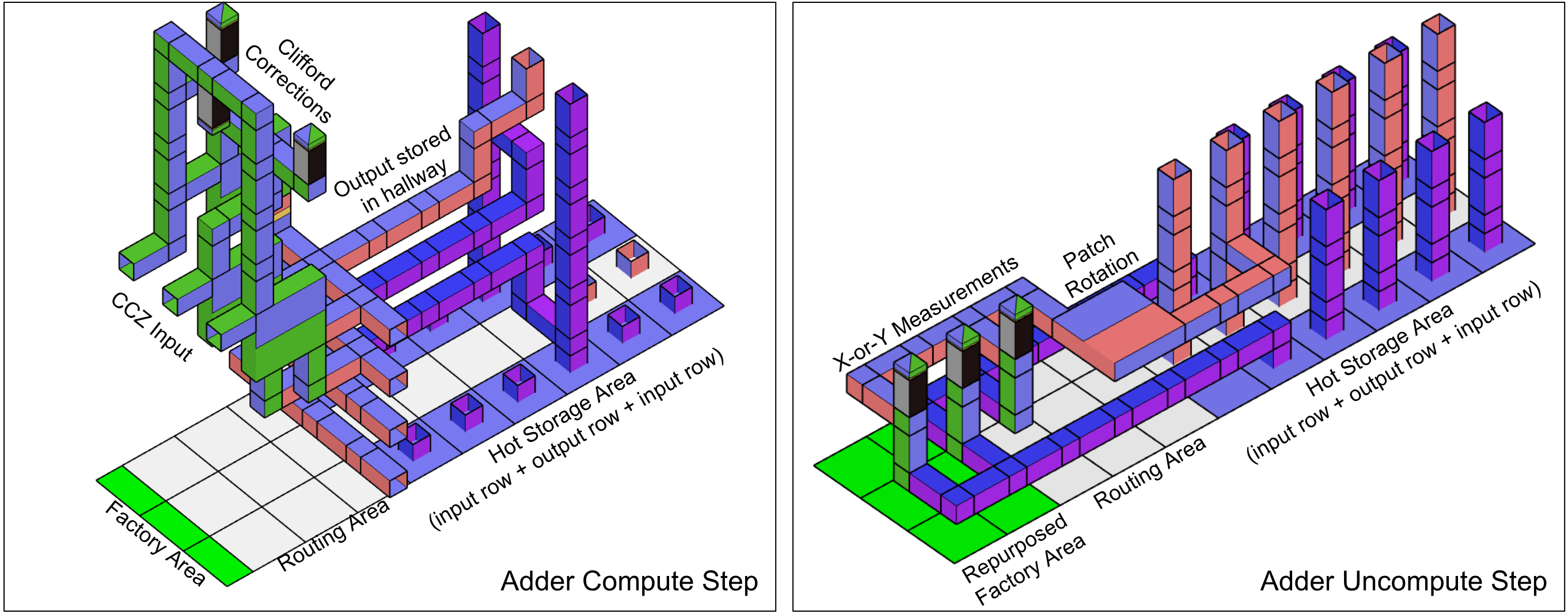}
    }
    \caption{
        Manual lattice surgery mock-up of adder compute and uncompute steps.
        Note that the steps tile horizontally, and Clifford corrections are being done after the operation (rather than in a reaction limited fashion).
        The uncomputation step assumes one of the inputs comes from a lookup, meaning its qubits can be removed by measurement based uncomputation with deferred phase corrections.
        Green highlights show CCZ-state-associated lattice surgery.
        Purple highlights show input-qubit-associated lattice surgery.
    }
    \label{fig:adder-step-mock-ups}
\end{figure}

\begin{figure}[ht]
    \centering
    \resizebox{\linewidth}{!}{
    \includegraphics{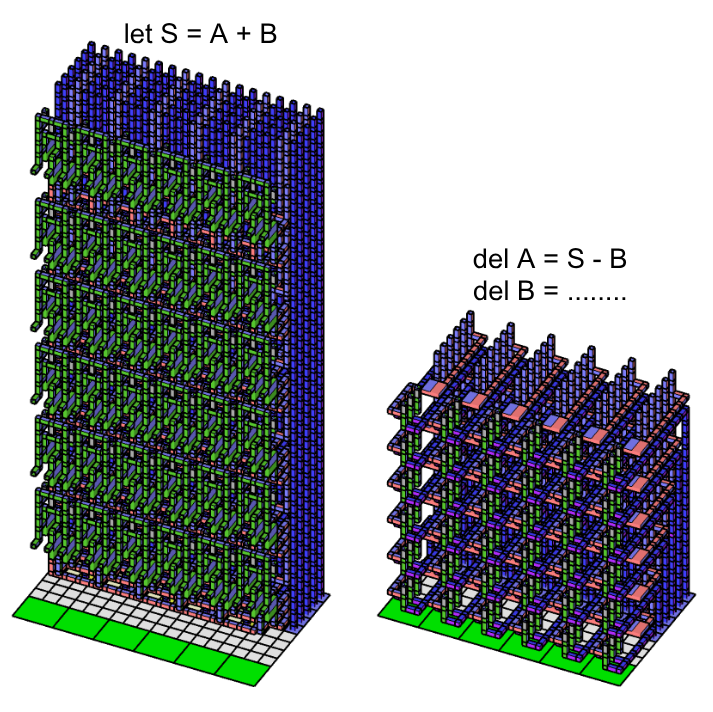}
    }
    \caption{
        Manual lattice surgery mock-up of a 35 qubit inplace addition.
        Only intended to give a qualitative sense of the high level layout; some details are omitted and the 3d model likely contains some minor mistakes.
        Uses many copies of \fig{adder-step-mock-ups}.
        The offset $B$ can be deleted in tandem because in context it comes from a lookup addressed by qubits stored elsewhere.
    }
    \label{fig:adder-mock-up}
\end{figure}

After the sum is computed, either of the inputs can be uncomputed using measurement based uncomputation as shown on the right hand side of \fig{adder-circuit}.
In context of the overall algorithm, usually one of the inputs is the previous value of an accumulator and the other input is the result of a looking up an offset.
So the previous accumulator value would be uncomputed as in \fig{adder-circuit}, and then the lookup value would be uncomputed using measurement based uncomputation creating phase corrections to be handled by a later phaseup operation, leaving behind only the new accumulator value.
Lattice surgery mock-ups of this process are shown in \fig{adder-step-mock-ups} and \fig{adder-mock-up}.

\FloatBarrier
\subsection{Phaseup Operation}
\label{app:phaseup-operation}

A ``phaseup'' is a quantum operation that negates the amplitude of computational basis states determined by a classical table of bits $T$.
A phaseup acts on an $n$ qubit address register and is driven by a $2^n$ bit table $T$.
Typically $n$ will be small, for example $n=6$, due to the exponential size of the table.
Phaseups appear when uncomputing table lookups with measurement based uncomputation~\cite{berry2019qubitization}.

\begin{equation}
\begin{aligned}
    \text{Phaseup}_T
    &= \rangesum{k}{2^n} \negif(T_k) |k\rangle \langle k|
\end{aligned}
\end{equation}

\begin{figure}[ht]
    \centering
    \resizebox{\linewidth}{!}{
        \includegraphics{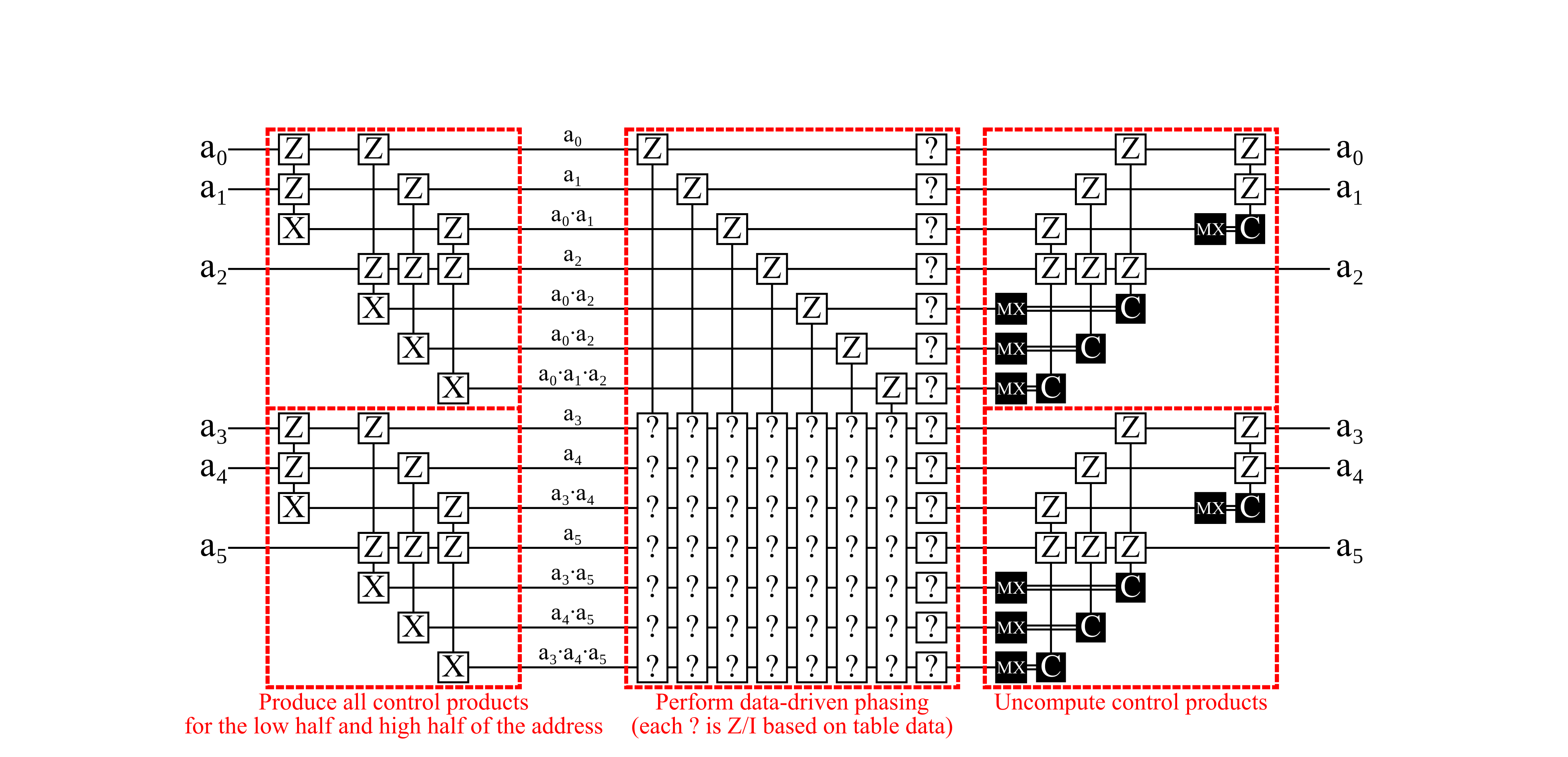}
    }
    \caption{
        Circuit diagram of a phaseup.
        On the left, the input register is split into two halves and each half is expanded into its power product using AND gates.
        In the middle, the desired phase flips are encoded into the presence/absence of Z and CZ gates.
        On the right, the power products are uncomputed using measurement based uncomputation.
        \href{
            https://algassert.com/quirk\#circuit=\%7b\%22cols\%22\%3a\%5b\%5b\%22\%7elbl0\%22,\%22\%7elbl1\%22,1,\%22\%7elbl2\%22,1,1,1,\%22\%7elbl3\%22,\%22\%7elbl4\%22,1,\%22\%7elbl5\%22\%5d,\%5b\%22\%e2\%80\%a2\%22,\%22\%e2\%80\%a2\%22,\%22X\%22\%5d,\%5b1,1,1,1,1,1,1,\%22\%e2\%80\%a2\%22,\%22\%e2\%80\%a2\%22,\%22X\%22\%5d,\%5b\%22\%e2\%80\%a2\%22,1,1,\%22\%e2\%80\%a2\%22,\%22X\%22\%5d,\%5b1,1,1,1,1,1,1,\%22\%e2\%80\%a2\%22,1,1,\%22\%e2\%80\%a2\%22,\%22X\%22\%5d,\%5b1,\%22\%e2\%80\%a2\%22,1,\%22\%e2\%80\%a2\%22,1,\%22X\%22\%5d,\%5b1,1,1,1,1,1,1,1,\%22\%e2\%80\%a2\%22,1,\%22\%e2\%80\%a2\%22,1,\%22X\%22\%5d,\%5b1,1,\%22\%e2\%80\%a2\%22,\%22\%e2\%80\%a2\%22,1,1,\%22X\%22\%5d,\%5b1,1,1,1,1,1,1,1,1,\%22\%e2\%80\%a2\%22,\%22\%e2\%80\%a2\%22,1,1,\%22X\%22\%5d,\%5b\%22zpar\%22,1,1,1,1,1,1,\%22Z\%22,\%22Z\%22,\%22Z\%22,\%22Z\%22,\%22Z\%22,\%22Z\%22,\%22Z\%22\%5d,\%5b1,\%22zpar\%22,1,1,1,1,1,\%22Z\%22,\%22Z\%22,\%22Z\%22,\%22Z\%22,\%22Z\%22,\%22Z\%22,\%22Z\%22\%5d,\%5b1,1,\%22zpar\%22,1,1,1,1,\%22Z\%22,\%22Z\%22,\%22Z\%22,\%22Z\%22,\%22Z\%22,\%22Z\%22,\%22Z\%22\%5d,\%5b1,1,1,\%22zpar\%22,1,1,1,\%22Z\%22,\%22Z\%22,\%22Z\%22,\%22Z\%22,\%22Z\%22,\%22Z\%22,\%22Z\%22\%5d,\%5b1,1,1,1,\%22zpar\%22,1,1,\%22Z\%22,\%22Z\%22,\%22Z\%22,\%22Z\%22,\%22Z\%22,\%22Z\%22,\%22Z\%22\%5d,\%5b1,1,1,1,1,\%22zpar\%22,1,\%22Z\%22,\%22Z\%22,\%22Z\%22,\%22Z\%22,\%22Z\%22,\%22Z\%22,\%22Z\%22\%5d,\%5b1,1,1,1,1,1,\%22zpar\%22,\%22Z\%22,\%22Z\%22,\%22Z\%22,\%22Z\%22,\%22Z\%22,\%22Z\%22,\%22Z\%22\%5d,\%5b\%22Z\%22,\%22Z\%22,\%22Z\%22,\%22Z\%22,\%22Z\%22,\%22Z\%22,\%22Z\%22,\%22Z\%22,\%22Z\%22,\%22Z\%22,\%22Z\%22,\%22Z\%22,\%22Z\%22,\%22Z\%22\%5d,\%5b1,1,1,1,\%22H\%22,\%22H\%22,\%22H\%22,1,1,1,1,\%22H\%22,\%22H\%22,\%22H\%22\%5d,\%5b1,1,1,1,\%22Measure\%22,\%22Measure\%22,\%22Measure\%22,1,1,1,1,\%22Measure\%22,\%22Measure\%22,\%22Measure\%22\%5d,\%5b1,1,\%22Z\%22,\%22\%e2\%80\%a2\%22,1,1,\%22\%e2\%80\%a2\%22\%5d,\%5b1,1,1,1,1,1,1,1,1,\%22Z\%22,\%22\%e2\%80\%a2\%22,1,1,\%22\%e2\%80\%a2\%22\%5d,\%5b1,\%22Z\%22,1,\%22\%e2\%80\%a2\%22,1,\%22\%e2\%80\%a2\%22\%5d,\%5b1,1,1,1,1,1,1,1,\%22Z\%22,1,\%22\%e2\%80\%a2\%22,1,\%22\%e2\%80\%a2\%22\%5d,\%5b\%22Z\%22,1,1,\%22\%e2\%80\%a2\%22,\%22\%e2\%80\%a2\%22\%5d,\%5b1,1,1,1,1,1,1,\%22Z\%22,1,1,\%22\%e2\%80\%a2\%22,\%22\%e2\%80\%a2\%22\%5d,\%5b1,1,\%22H\%22,1,1,1,1,1,1,\%22H\%22\%5d,\%5b1,1,\%22Measure\%22,1,1,1,1,1,1,\%22Measure\%22\%5d,\%5b\%22Z\%22,\%22\%e2\%80\%a2\%22,\%22\%e2\%80\%a2\%22\%5d,\%5b\%22\%7elbl0\%22,\%22\%7elbl1\%22,1,\%22\%7elbl2\%22,1,1,1,\%22Z\%22,\%22\%e2\%80\%a2\%22,\%22\%e2\%80\%a2\%22\%5d,\%5b1,1,1,1,1,1,1,\%22\%7elbl3\%22,\%22\%7elbl4\%22,1,\%22\%7elbl5\%22\%5d\%5d,\%22gates\%22\%3a\%5b\%7b\%22id\%22\%3a\%22\%7elbl0\%22,\%22name\%22\%3a\%22a\%e2\%82\%80\%22,\%22matrix\%22\%3a\%22\%7b\%7b1,0\%7d,\%7b0,1\%7d\%7d\%22\%7d,\%7b\%22id\%22\%3a\%22\%7elbl1\%22,\%22name\%22\%3a\%22a\%e2\%82\%81\%22,\%22matrix\%22\%3a\%22\%7b\%7b1,0\%7d,\%7b0,1\%7d\%7d\%22\%7d,\%7b\%22id\%22\%3a\%22\%7elbl2\%22,\%22name\%22\%3a\%22a\%e2\%82\%82\%22,\%22matrix\%22\%3a\%22\%7b\%7b1,0\%7d,\%7b0,1\%7d\%7d\%22\%7d,\%7b\%22id\%22\%3a\%22\%7elbl3\%22,\%22name\%22\%3a\%22a\%e2\%82\%83\%22,\%22matrix\%22\%3a\%22\%7b\%7b1,0\%7d,\%7b0,1\%7d\%7d\%22\%7d,\%7b\%22id\%22\%3a\%22\%7elbl4\%22,\%22name\%22\%3a\%22a\%e2\%82\%84\%22,\%22matrix\%22\%3a\%22\%7b\%7b1,0\%7d,\%7b0,1\%7d\%7d\%22\%7d,\%7b\%22id\%22\%3a\%22\%7elbl5\%22,\%22name\%22\%3a\%22a\%e2\%82\%85\%22,\%22matrix\%22\%3a\%22\%7b\%7b1,0\%7d,\%7b0,1\%7d\%7d\%22\%7d\%5d\%7d
        }{Click here to open this circuit in Quirk.}
    }
    \label{fig:phaseup-circuit-diagram}
\end{figure}

For this paper, I imagined phaseups to be implemented by splitting the register into a low half $L$ (with the $n // 2$ least significant qubits) and a high half $H$ (with the $n - n//2$ most significant qubits).
This effectively reshapes $T$ into a $2^{\len L} \times 2^{\len H}$ matrix, with rows/columns indexed by $H, L$.
Each half-register would then be expanded into its ``power product'': a register storing every possible product that can be formed out of the original register's qubits.
For example, if $L$ was a three-qubit register storing the computational basis state $|c, b, a\rangle$ then its power product $L^\ast$ is the computational basis state $|cba, cb, ca, c, ab, b, a, 1\rangle$.
Power products can be computed with a series of AND gates (for example, see the left side of \fig{phaseup-circuit-diagram}).
Once the power products are computed, they can implement the phaseup operation using a series of ``masked phase flips''.
A masked phase flip is a multi-controlled Z gate, with the involved qubits determined by the positions of 1s within a binary mask $m$:

\begin{equation}
    \text{MaskedPhaseFlip}_n(m) = \rangesum{k}{2^n} \negif(k\;\&\;m = m) |k\rangle \langle k|
\end{equation}

To implement the desired phaseup in terms of masked phase flips, it's necessary to determine the values of the masks.
The data in $T$ lists whether or not to phase flip each possible value of the register pair $(H, L)$.
We instead want an alternate form $\overline{T}$ that lists whether or not to include/exclude each possible masked phase flip (i.e. the coefficients of the phaseup's EXOR polynomial~\cite{mishchenko2001exor}).
Conveniently, this conversion corresponds to a matrix multiplication:

\begin{equation}
    \overline{T} = \begin{bmatrix}1 & 0 \\ 1 & 1\end{bmatrix}^{\otimes \len H} \cdot T \cdot \begin{bmatrix}1 & 1 \\ 0 & 1\end{bmatrix}^{\otimes \len L} \pmod{2}
\end{equation}

\begin{equation}
\begin{aligned}
    \text{Phaseup}_T
    &=
    \prod_{l=0}^{(\len L^\ast)-1}
    \prod_{h=0}^{(\len H^\ast)-1}
    \text{CZ}(H^\ast_h, L^\ast_l)^{\overline{T}_{h,l}}
\end{aligned}
\end{equation}

In \fig{phaseup-circuit-diagram} I show a circuit diagram of a phaseup implemented this way, including the data-driven CZ gates as well as the computation and uncomputation of the power products.
Note that this circuit has optimized out the two trivial qubits $L^*_0 = H^\ast_0 = |1\rangle$, resulting in some CZ gates becoming Z gates and one CZ gate becoming an irrelevant global phase (not shown).
Also note that \fig{phaseup-circuit-diagram} has grouped the many individual CZ gates into a few large multi-target CZ gates.
This is beneficial because in lattice surgery the marginal cost of targeting a larger Pauli product is notably lower than the marginal cost of doing another gate~\cite{fowler2018latticesurgery,litinski2018}.

I apply one key additional optimization beyond what's shown \fig{phaseup-circuit-diagram}.
I allow the AND computations that appear during the computation of the power product to ``wander''.
That is to say, I perform AND gates by gate teleportation but don't correct the teleportation.
An AND gate computes $c = a \cdot b$, whereas a wandering AND gate computes $c = (a \oplus x) \cdot (b \oplus y)$ for measured random values $x$ and $y$.
Normally these $x$ and $y$ values are removed with corrective CNOT gates, performed on the quantum computer.
Instead, I account for the CNOT gates by computing a correction matrix $C$ to be multiplied into $\overline{T}$.
($C$ also affects the phase corrections performed during the uncomputation of the power products.)
In other words, the CNOT corrections created by the wandering AND gates are handled by rewrites in the classical control system instead of by extra quantum gates.
This is analogous to how \cite{litinski2018} skips performing Clifford gates by folding their effects into the Pauli product operators targeted by $T$ gates and measurement operations.

Because wandering AND gates are used when computing the power products, computing the power products has constant reaction depth.
Instead of needing to separately correct each layer of AND gates, the corrections merge into a single massive change to the set of performed Z and CZ gates.
Only the uncomputation needs to be corrected layer by layer.
So, interestingly, the overall reaction depth of the phaseup is $n/2 \pm O(1)$ instead of $n \pm O(1)$.

\begin{figure}[ht]
    \centering
    \resizebox{\linewidth}{!}{
    \includegraphics{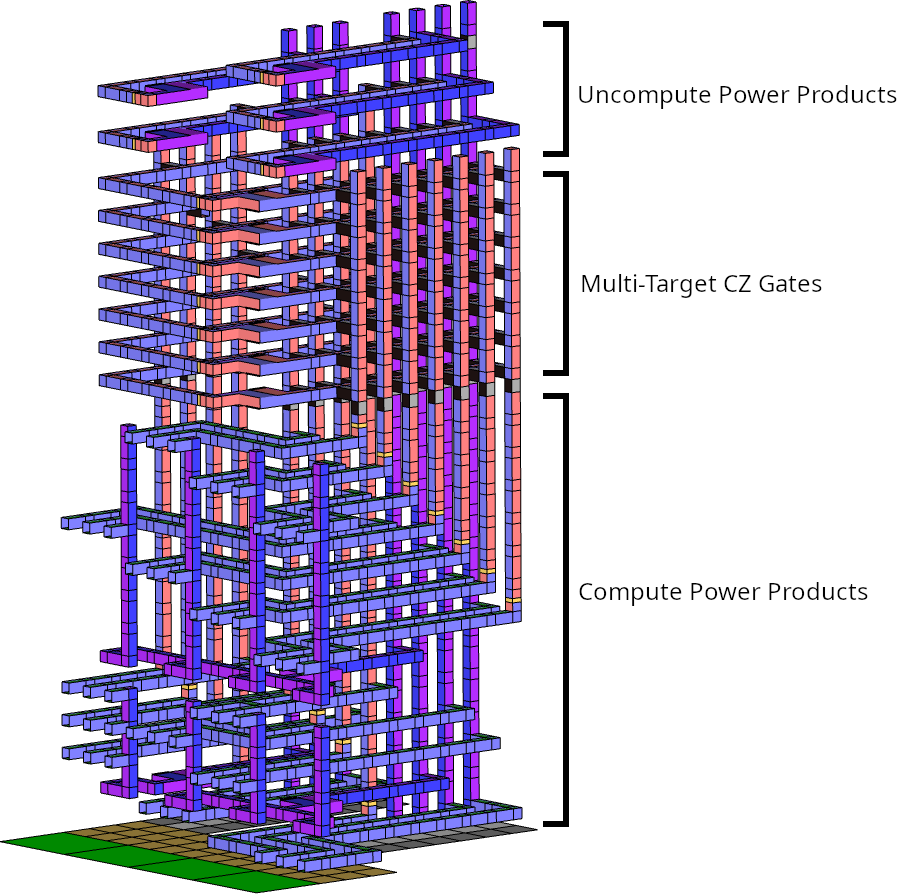}
    }
    \caption{
        Manual mock-up of lattice surgery performing a phaseup operation.
        Only intended to give a qualitative sense of the high level layout; some details are omitted and the 3d model likely contains some minor mistakes.
        Address qubits are accented with purple, CCZ qubits are accented with green, output qubits are accented with red, and delays for control system reaction time are accented with black and white.
    }
    \label{fig:phaseup-mock-up}
\end{figure}

Overall, a phaseup can be done with $2\sqrt{2^n}$ AND gates, $2\sqrt{2^n}$ workspace qubits, and $\sqrt{2^n}$ multi-target CZ gates.
This is reminiscent of the $O(\sqrt{n})$ costs of select-swap lookups~\cite{low2018qrom}, which similarly starts by dividing the input register into two halves.
See \fig{phaseup-mock-up} for a lattice surgery mock-up of a phaseup operation implemented in this way.

\subsection{Lookup Operation}
\label{app:lookup-operation}

A lookup is an operation that initializes $w$ output qubits using values from a classical table storing $2^n$ $w$-bit integers, indexed by an $n$ qubit quantum address:

\begin{equation}
    \text{Lookup}_T = \rangesum{k}{2^n} \Big( |T_{k}\rangle \otimes |k\rangle \Big) \langle k|
\end{equation}

For this paper, I imagined implementing lookups similar to phaseups.
First, split the address register into two halves and compute the power product of each half.
Second, perform multi-target Toffoli gates (one for each pair of control qubits from the two half registers) to initialize the output qubits.
As in the phaseup, the AND gates in the power product computation are allowed to wander.
But now the multi-target Toffoli gates are also allowed to wander.
This creates dependencies between the different Toffoli gates, where the teleportation outcome of one can change the qubits that must be targeted by another.
These changes can be accounted for by the classical control system, but the Toffolis must be carefully ordered to avoid having to redo work and to avoid incurring reaction delays.

After the Toffolis finish, the lookup is completed by measuring all ancillary qubits in the X basis.
This creates an enormous number of phase corrections, but every one of these corrections corresponds to phase flipping some subset of the address values.
Therefore all these corrections can be merged into a phaseup, which can be deferred and merged into the phaseup that appears in the uncomputation of the lookup later on.

\begin{figure}[ht]
    \centering
    \resizebox{0.8\linewidth}{!}{
    \includegraphics{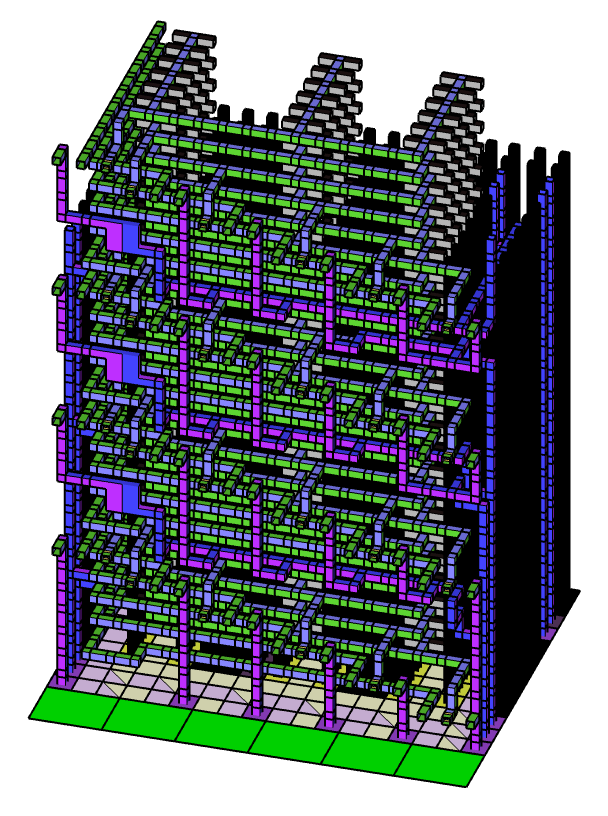}
    }
    \caption{
        Manual mock-up of lattice surgery for a lookup operation.
        Only intended to give a qualitative sense of the high level layout; some details are omitted and the 3d model likely contains some minor mistakes.
        21 multi-target CX gates are included in the diagram, out of the 63 that would be performed by a lookup with 6 address qubits.
        The computation of the two power products isn't shown.
        Green-accented pipes are CCZ qubits, purple-accented pipes are power product qubits, and black pipes are immobile qubits.
        One of the CCZ qubits pauses momentarily when leaving the factory, before rotating to become the target of a Toffoli state, to give time for the control system to update its targets in response to previous gate teleportations.
        One of the power products is keeping its qubits still, to be reached by one of the qubits of the CCZ state during each multi-target Toffoli.
        The other power product is proactively moving its qubits in front of the magic state factories, to intercept one qubit from each CCZ state.
    }
    \label{fig:lookup-mock-up}
\end{figure}

I show a mock-up for the lattice surgery of a lookup in \fig{lookup-mock-up}.

\FloatBarrier
\subsection{Frequency Basis Measurement}

\label{sec:frequency-basis-measurement}

\begin{figure}
    \centering
    \resizebox{\linewidth}{!}{
    \includegraphics{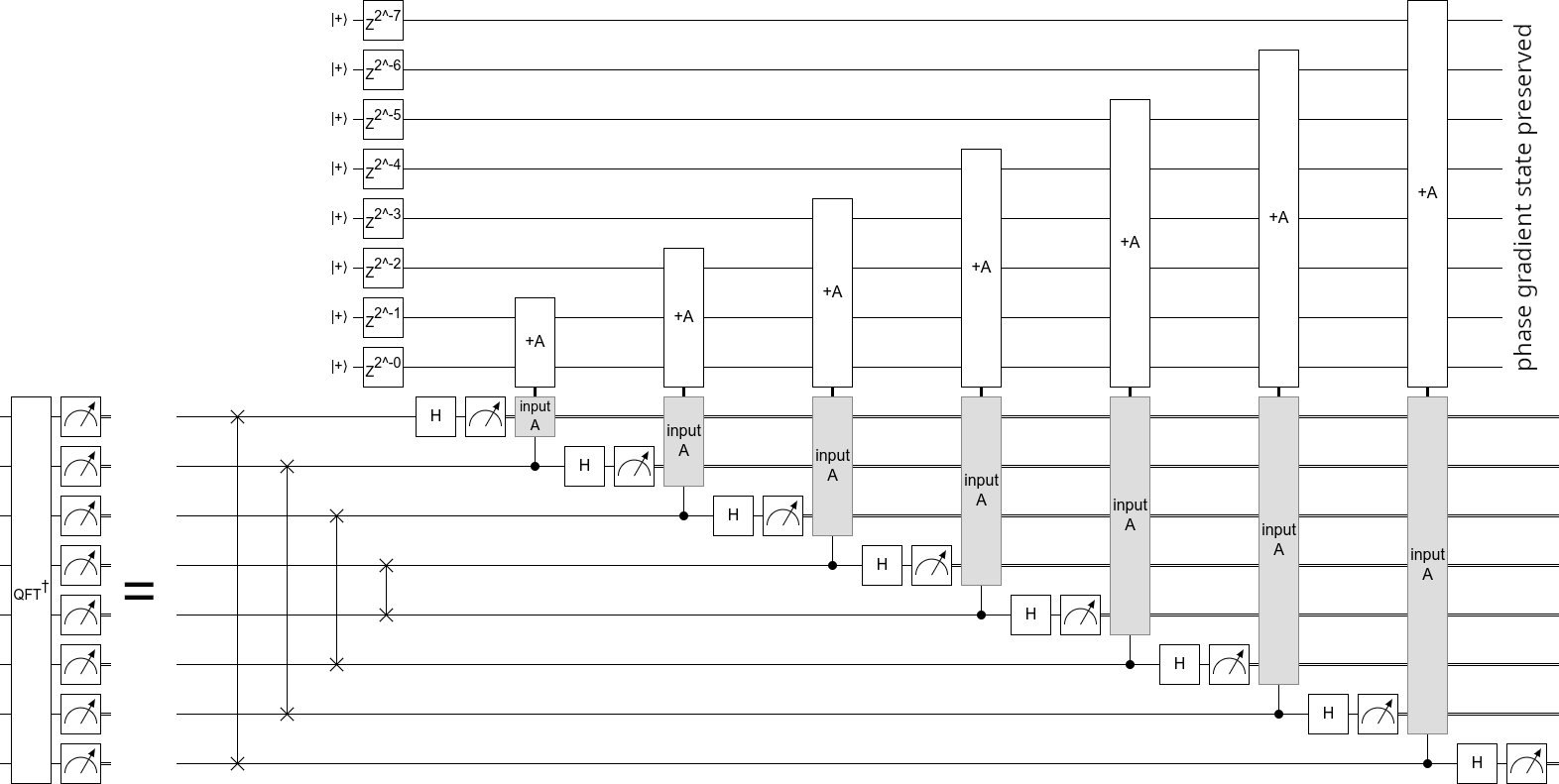}
    }
    \caption{
        A frequency basis measurement implemented with adaptive phase gates~\cite{Parker2000,Mosca1999} achieved by kickback from controlled addition of a classical constant into a phase gradient state~\cite{Kitaev2002,gidney2018addition}.
        The phase gradient state can be truncated, resulting in an approximate QFT~\cite{Nam2020}.
        \href{
            https://algassert.com/quirk\#circuit=\%7B\%22cols\%22\%3A\%5B\%5B\%7B\%22id\%22\%3A\%22Z\%5Eft\%22\%2C\%22arg\%22\%3A\%222\%5E-7\%22\%7D\%2C\%7B\%22id\%22\%3A\%22Z\%5Eft\%22\%2C\%22arg\%22\%3A\%222\%5E-6\%22\%7D\%2C\%7B\%22id\%22\%3A\%22Z\%5Eft\%22\%2C\%22arg\%22\%3A\%222\%5E-5\%22\%7D\%2C\%7B\%22id\%22\%3A\%22Z\%5Eft\%22\%2C\%22arg\%22\%3A\%222\%5E-4\%22\%7D\%2C\%7B\%22id\%22\%3A\%22Z\%5Eft\%22\%2C\%22arg\%22\%3A\%222\%5E-3\%22\%7D\%2C\%7B\%22id\%22\%3A\%22Z\%5Eft\%22\%2C\%22arg\%22\%3A\%222\%5E-2\%22\%7D\%2C\%7B\%22id\%22\%3A\%22Z\%5Eft\%22\%2C\%22arg\%22\%3A\%222\%5E-1\%22\%7D\%2C\%7B\%22id\%22\%3A\%22Z\%5Eft\%22\%2C\%22arg\%22\%3A\%222\%5E-0\%22\%7D\%5D\%2C\%5B1\%2C1\%2C1\%2C1\%2C1\%2C1\%2C1\%2C1\%2C\%22Swap\%22\%2C1\%2C1\%2C1\%2C1\%2C1\%2C1\%2C\%22Swap\%22\%5D\%2C\%5B1\%2C1\%2C1\%2C1\%2C1\%2C1\%2C1\%2C1\%2C1\%2C\%22Swap\%22\%2C1\%2C1\%2C1\%2C1\%2C\%22Swap\%22\%5D\%2C\%5B1\%2C1\%2C1\%2C1\%2C1\%2C1\%2C1\%2C1\%2C1\%2C1\%2C\%22Swap\%22\%2C1\%2C1\%2C\%22Swap\%22\%5D\%2C\%5B1\%2C1\%2C1\%2C1\%2C1\%2C1\%2C1\%2C1\%2C1\%2C1\%2C1\%2C\%22Swap\%22\%2C\%22Swap\%22\%5D\%2C\%5B1\%2C1\%2C1\%2C1\%2C1\%2C1\%2C1\%2C1\%2C\%22H\%22\%5D\%2C\%5B1\%2C1\%2C1\%2C1\%2C1\%2C1\%2C1\%2C1\%2C\%22Measure\%22\%5D\%2C\%5B1\%2C1\%2C1\%2C1\%2C1\%2C1\%2C\%22\%2B\%3DA2\%22\%2C1\%2C\%22inputA1\%22\%2C\%22\%E2\%80\%A2\%22\%5D\%2C\%5B1\%2C1\%2C1\%2C1\%2C1\%2C1\%2C1\%2C1\%2C1\%2C\%22H\%22\%5D\%2C\%5B1\%2C1\%2C1\%2C1\%2C1\%2C1\%2C1\%2C1\%2C1\%2C\%22Measure\%22\%5D\%2C\%5B1\%2C1\%2C1\%2C1\%2C1\%2C\%22\%2B\%3DA3\%22\%2C1\%2C1\%2C\%22inputA2\%22\%2C1\%2C\%22\%E2\%80\%A2\%22\%5D\%2C\%5B1\%2C1\%2C1\%2C1\%2C1\%2C1\%2C1\%2C1\%2C1\%2C1\%2C\%22H\%22\%5D\%2C\%5B1\%2C1\%2C1\%2C1\%2C1\%2C1\%2C1\%2C1\%2C1\%2C1\%2C\%22Measure\%22\%5D\%2C\%5B1\%2C1\%2C1\%2C1\%2C\%22\%2B\%3DA4\%22\%2C1\%2C1\%2C1\%2C\%22inputA3\%22\%2C1\%2C1\%2C\%22\%E2\%80\%A2\%22\%5D\%2C\%5B1\%2C1\%2C1\%2C1\%2C1\%2C1\%2C1\%2C1\%2C1\%2C1\%2C1\%2C\%22H\%22\%5D\%2C\%5B1\%2C1\%2C1\%2C1\%2C1\%2C1\%2C1\%2C1\%2C1\%2C1\%2C1\%2C\%22Measure\%22\%5D\%2C\%5B1\%2C1\%2C1\%2C\%22\%2B\%3DA5\%22\%2C1\%2C1\%2C1\%2C1\%2C\%22inputA4\%22\%2C1\%2C1\%2C1\%2C\%22\%E2\%80\%A2\%22\%5D\%2C\%5B1\%2C1\%2C1\%2C1\%2C1\%2C1\%2C1\%2C1\%2C1\%2C1\%2C1\%2C1\%2C\%22H\%22\%5D\%2C\%5B1\%2C1\%2C1\%2C1\%2C1\%2C1\%2C1\%2C1\%2C1\%2C1\%2C1\%2C1\%2C\%22Measure\%22\%5D\%2C\%5B1\%2C1\%2C\%22\%2B\%3DA6\%22\%2C1\%2C1\%2C1\%2C1\%2C1\%2C\%22inputA5\%22\%2C1\%2C1\%2C1\%2C1\%2C\%22\%E2\%80\%A2\%22\%5D\%2C\%5B1\%2C1\%2C1\%2C1\%2C1\%2C1\%2C1\%2C1\%2C1\%2C1\%2C1\%2C1\%2C1\%2C\%22H\%22\%5D\%2C\%5B1\%2C1\%2C1\%2C1\%2C1\%2C1\%2C1\%2C1\%2C1\%2C1\%2C1\%2C1\%2C1\%2C\%22Measure\%22\%5D\%2C\%5B1\%2C\%22\%2B\%3DA7\%22\%2C1\%2C1\%2C1\%2C1\%2C1\%2C1\%2C\%22inputA6\%22\%2C1\%2C1\%2C1\%2C1\%2C1\%2C\%22\%E2\%80\%A2\%22\%5D\%2C\%5B1\%2C1\%2C1\%2C1\%2C1\%2C1\%2C1\%2C1\%2C1\%2C1\%2C1\%2C1\%2C1\%2C1\%2C\%22H\%22\%5D\%2C\%5B1\%2C1\%2C1\%2C1\%2C1\%2C1\%2C1\%2C1\%2C1\%2C1\%2C1\%2C1\%2C1\%2C1\%2C\%22Measure\%22\%5D\%2C\%5B\%22\%2B\%3DA8\%22\%2C1\%2C1\%2C1\%2C1\%2C1\%2C1\%2C1\%2C\%22inputA7\%22\%2C1\%2C1\%2C1\%2C1\%2C1\%2C1\%2C\%22\%E2\%80\%A2\%22\%5D\%2C\%5B1\%2C1\%2C1\%2C1\%2C1\%2C1\%2C1\%2C1\%2C1\%2C1\%2C1\%2C1\%2C1\%2C1\%2C1\%2C\%22H\%22\%5D\%2C\%5B1\%2C1\%2C1\%2C1\%2C1\%2C1\%2C1\%2C1\%2C1\%2C1\%2C1\%2C1\%2C1\%2C1\%2C1\%2C\%22Measure\%22\%5D\%2C\%5B1\%2C\%22\%E2\%80\%A6\%22\%5D\%5D\%2C\%22init\%22\%3A\%5B\%22\%2B\%22\%2C\%22\%2B\%22\%2C\%22\%2B\%22\%2C\%22\%2B\%22\%2C\%22\%2B\%22\%2C\%22\%2B\%22\%2C\%22\%2B\%22\%2C\%22\%2B\%22\%5D\%7D
        }{Click here to open this circuit in Quirk.}
    }
    \label{fig:qft-measure}
    \vspace{20mm}
    \centering
    \resizebox{\linewidth}{!}{
    \includegraphics{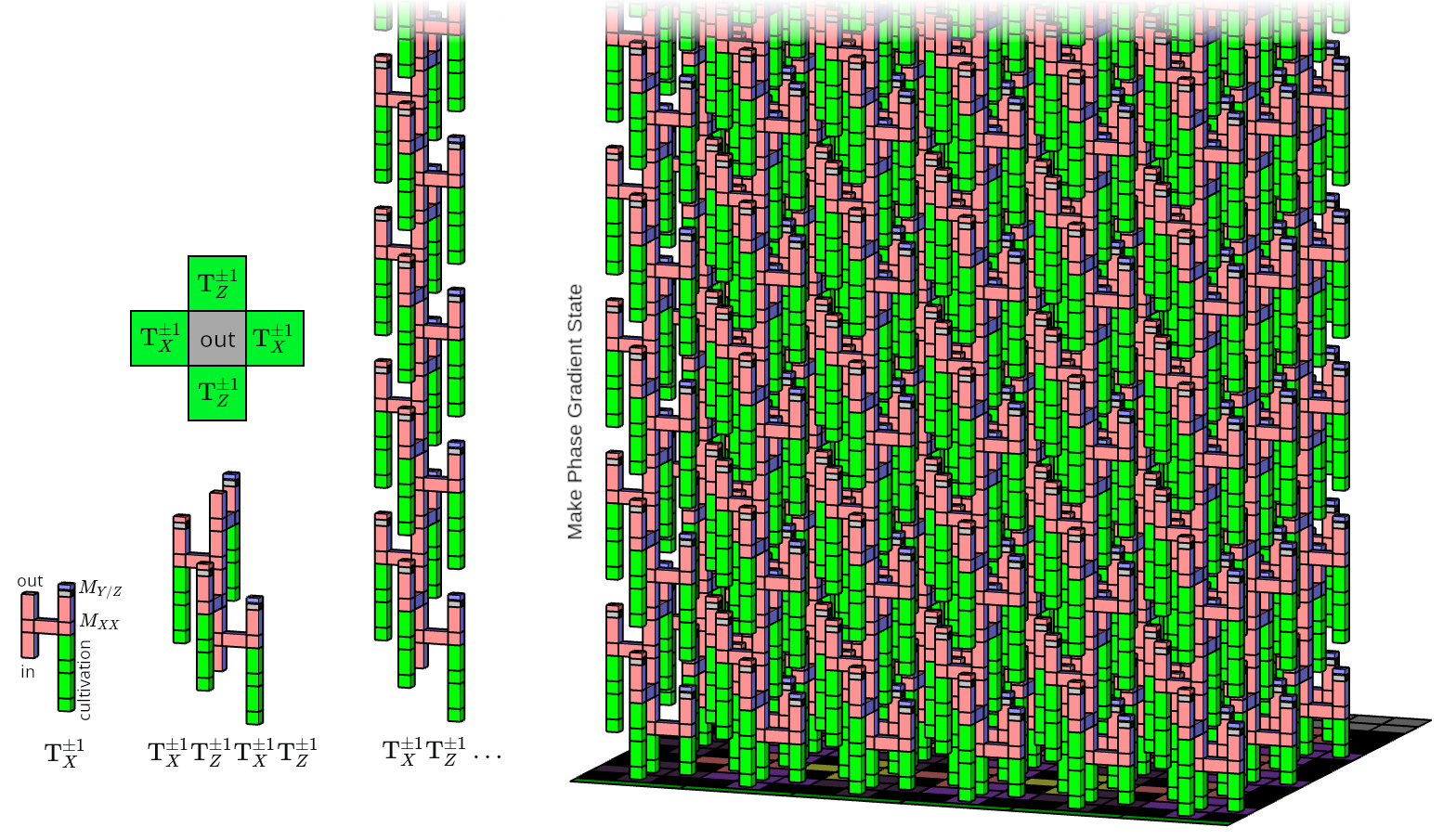}
    }
    \caption{
        Manual mock-up of lattice surgery preparing a phase gradient state.
        Left: applying a $T_X^{\pm 1}$ gate by magic state cultivation and teleportation.
        The sign of the exponent is controlled by picking the basis of the $M_{Y/Z}$ measurement, accounting for the result of the $M_{XX}$ pair measurement~\cite{litinski2018,gidney2019autoccz}.
        Left middle: a central qubit can be surrounded with $T_X^{\pm 1}$ and $T_Z^{\pm 1}$ gates taking turns operating on it.
        Middle: this process can be continued as long as needed.
        Right: many phase gradient qubits being prepared in parallel.
    }
    \label{fig:phase-gradient-lattice-surgery}
\end{figure}

A frequency basis measurement is implemented by performing an inverse Quantum Fourier transform (QFT) and then measuring in the computational basis.
Because of the presence of the measurements, the deferred measurement principle can be used to merge the quadratically many controlled phase gates that normally appear in a QFT circuit into $n-1$ phase gates with adaptively determined phase angles~\cite{Parker2000,Mosca1999} (see \fig{qft-measure}).

To implement phase gates, I use kickback from additions into a phase gradient state~\cite{Kitaev2002,gidneygradqft2016,gidney2018addition,Nam2020}.
A phase gradient state is a $g$ qubit state where qubit $k$ is in the state $Z^{2^{-k}}|+\rangle$.
Equivalently, it's a uniform superposition where the amplitude for the computational basis state $|k\rangle$ has been phased by $k/2^n$ turns:

\begin{equation}
    |\text{GRAD}_{g}\rangle
    = 2^{-g/2} \rangesum{k}{2^g} |k\rangle e^{i 2\pi k/2^g}
    = \bigotimes_{k=0}^{g-1} Z^{2^{-k}} |+\rangle
\end{equation}

If a classical constant $C$ is added into a $g$ qubit phase gradient state, controlled by a qubit $q$, then phase kickback will phase $q$ by $-C/2^g$ turns while leaving the phase gradient state unchanged.
$C$ is chosen by rounding the desired angle $\theta$ to the nearest multiple of $2 \pi / 2^g$.

Rounding $\theta$ introduces a per-rotation error.
The maximum over/under rotation is $\pi / 2^g$ radians.
The chance of a shot failure due to this rounding, across the $n-1$ rotations done by the frequency basis measurement, is at most $n \frac{\pi}{2^g}$.

Another source of error is the infidelity of the phase gradient state, due to approximate preparation.
This error differs from the rounding error in that it only applies once per state, instead of once per use.
The phase gradient state is an eigenstate of addition and, once prepared, is only used as the target of additions.
Therefore the prepared state could be measured in the eigenbasis of addition without changing the behavior of the algorithm, projecting the prepared state into a perfect state with probability equal to its fidelity.
So the phase gradient preparation contributes a shot failure probability no larger than the infidelity of the state, regardless of how many times the state is used.

To prepare phase gradient qubits, I use single qubit Clifford+T sequences~\cite{selinger2014gridsynth}.
I convert the sequences into a form that begins with a Pauli basis initialization, ends with a Clifford rotation, and otherwise alternates between performing $T_X^{\pm 1}$ and $T_Z^{\pm 1}$ where $T_Z = Z^{1/4}$ and $T_X = X^{1/4}$.
In terms of lattice surgery, I imagine executing the sequence with the help of four neighboring patches.
Each neighbor will repeatedly cultivate a T state, undergo a parity measurement against the target patch, and then undergo an adaptively chosen measurement.
When a $T_X^{\pm 1}$ gate is needed, an $M_{XX}$ measurement is performed between the target patch and one of its $X$-boundary neighbors that has a T state ready.
This measurement reveals whether or not the target was rotated by $+45^\circ$ or $-45^\circ$ around the $X$ axis.
If it rotated in the correct direction, the neighboring patch is measured in the $Z$ basis.
Otherwise it's measured in the Y basis, correcting the direction of the rotation~\cite{litinski2018} (up to classically tracked Paulis).
The same story occurs for $T_Z^{\pm 1}$ gates, but with X and Z swapped.
This strategy results in lattice surgery operations spiraling around the target patch, with the neighbors taking turns contributing T states (see \fig{phase-gradient-lattice-surgery}).

Concretely, I would use the gate sequences shown in \tbl{gradient-decompositions} to prepare the phase gradient state's qubits.
The first 11 qubits of the state are prepared using a total of 159 T gates.
All further qubits, up to the desired length $g$, are approximated to high fidelity by $|+\rangle$ states.
This preparation has an infidelity of $5 \cdot 10^{-6}$, assuming perfect gates.
If the T states powering the gates are cultivated to an infidelity of $10^{-7}/4$, then inaccurate T gates will introduce an additional infidelity of $4 \cdot 10^{-6}$.
So, in total, preparing the phase gradient state in this way contributes a less than $10^{-5}$ chance of algorithm failure.
This chance could be reduced, by distilling better T states and using the more accurate gate sequences from \tbl{gradient-decompositions-accurate}.
But ultimately the error costs and operation costs of the frequency basis measurement are completely negligible compared to other costs in the paper, so for the purposes of cost estimates I simply ignore it.

\begin{table}
    \centering
    \input{assets/man/tbl-gradient-decomposition}
    \caption{
        Example decompositions of phase gradient qubit initializations, with a per-sequence infidelity target of $10^{-6}$.
        Found using a meet-in-the-middle search.
        Each qubit is prepared with a Pauli basis initialization, then alternating $T_X^{\pm 1}$ and $T_Z^{\pm 1}$ gates, then a Clifford rotation.
        The string of signs specifies the sign of each $T_X^{\pm 1}$ and $T_Z^{\pm 1}$ gate in sequence, with the leftmost sign corresponding to the first $T_X^{\pm 1}$ gate.
    }
    \vspace{10mm}
    \label{tbl:gradient-decompositions}

\input{assets/man/tbl-gradient-decomposition-accurate}

    \caption{
        Same as \tbl{gradient-decompositions} but with a per-qubit infidelity target of $10^{-15}$.
    }
    \label{tbl:gradient-decompositions-accurate}
\end{table}

\end{document}

%% file: assets/man/tbl-notation.tex
    \centering
    \resizebox{\linewidth}{!}{
    \begin{tabular}{|c|c|l|}
        \hline
        \textbf{Notation} & \textbf{Equivalent To} & \textbf{Description} \\
        \hline
        $\len a$ & $\lceil \log_2 \max(1, a) \rceil$ & Min bits needed to store an integer in the range $[0, a)$. 
        \\
        \hline
        $\len q$ & - & Number of qubits in a quantum register $q$. 
        \\
        \hline
        $a // b$ & $\lfloor a/b \rfloor$ & Floored division of $a$ by $b$.
        \\
        \hline
        $a \bmod b$ & $a - b \cdot (a // b)$ & Remainder of $a$ divided by $b$, canonicalized into $[0, b)$.
        \\
        \hline
        $a \bmod b \bmod c$ & $(a \bmod b) \bmod c$ & $\bmod$ is left-associative.
        \\
        \hline
        $a \gg b$ & $a//2^b$ & Right shift of $a$ by $b$ bits.
        \\
        \hline
        $a \ll b$ & $a2^b$ & Left shift of $a$ by $b$ bits.
        \\
        \hline
        $|a:b:c\rangle$ & $\displaystyle\rangesum{k}{\lceil (b-a)/c \rceil} \frac{|kc + a\rangle}{\sqrt{\lceil (b-a)/c \rceil}}$ & Uniform superposition, from $a$ to $b-1$, stepping by $c$.
        \\
        \hline
        $\text{GRAD}_{a}^b$ & $\displaystyle\rangesum{k}{a} |k\rangle\langle k| e^{ik 2\pi b/a}$ & Phase gradient modulo $a$, scaled by $b$.
        \\
        \hline
        $\text{QFT}_{a}$ & $\displaystyle\frac{1}{\sqrt{a}}\rangesum{k}{a}\rangesum{j}{a} |j\rangle\langle k| e^{ijk 2\pi /a}$ & Quantum Fourier Transform modulo $a$.
        \\
        \hline
        $\Delta_N(a)$
        & $\min(a \bmod N, -a \bmod N) / N$
        & Modular deviation of $a$, relative to $N$.
        \\
        \hline
        $\text{int}(a)$
        &
            $\begin{cases} a &\rightarrow 1\\\lnot a &\rightarrow 0 \end{cases}$
        & Indicator function for a boolean expression.
        \\
        \hline
        $\text{negif}(a)$
        &
            $(-1)^{\text{int}(a)}$
        & Phase flip indicator function for a boolean expression.
        \\
        \hline
        $a_b$
        &
            $(a \gg b) \mod 2$
        & Access bit (or qubit) $b$ within integer (or quint) $a$.
        \\
        \hline
        $a \;\&\; b$
        & $\sum_k a_k b_k \ll k$
        & Bitwise AND of $a$ and $b$.
        \\
        \hline
        $\text{MultiplicativeInverse}_a(b)$
        & $b^{-1} \pmod{a}$
        & Multiplicative inverse of $b$ modulo $a$.
        \\
        \hline
    \end{tabular}
    }
    \caption{
        Various bits of mostly-Python-inspired notation used in this paper.
    }

%% file: assets/gen/logical-cost-table.tex
\begin{tabular}{|c|c|c|c|c|c|c|c|c|c|c|c|}
\hline
\text{$n$} & \text{$s$} & \text{$\ell$} & \text{$w_1$} & \text{$w_3$} & \text{$w_4$} & \text{$f$} & \text{$m$} & \text{$P_{\text{deviant}}$} & \text{E(shots)} & \text{Toffolis} & \text{Qubits}
\\
\hline
\text{1024} & \text{8} & \text{18} & \text{6} & \text{3} & \text{6} & \text{28} & \text{640} & \text{2.87\%} & \text{9.4} & \text{1.1e+09} & \text{742}
\\
\text{1536} & \text{8} & \text{21} & \text{6} & \text{3} & \text{5} & \text{31} & \text{960} & \text{1.83\%} & \text{9.3} & \text{3.1e+09} & \text{1074}
\\
\text{2048} & \text{8} & \text{21} & \text{6} & \text{3} & \text{5} & \text{33} & \text{1280} & \text{1.25\%} & \text{9.2} & \text{6.5e+09} & \text{1399}
\\
\text{3072} & \text{8} & \text{21} & \text{6} & \text{3} & \text{5} & \text{35} & \text{1920} & \text{0.91\%} & \text{9.2} & \text{1.9e+10} & \text{2043}
\\
\text{4096} & \text{8} & \text{24} & \text{6} & \text{3} & \text{5} & \text{36} & \text{2560} & \text{0.80\%} & \text{9.2} & \text{4.0e+10} & \text{2692}
\\
\text{6144} & \text{8} & \text{24} & \text{6} & \text{3} & \text{5} & \text{39} & \text{3840} & \text{0.42\%} & \text{9.1} & \text{1.2e+11} & \text{3978}
\\
\text{8192} & \text{8} & \text{24} & \text{6} & \text{3} & \text{5} & \text{40} & \text{5120} & \text{0.40\%} & \text{9.1} & \text{2.7e+11} & \text{5261}
\\
\hline
\end{tabular}

%% file: assets/man/tbl-gradient-decomposition.tex
    \resizebox{\linewidth}{!}{
    \begin{tabular}{|c|c|l|c|c|c|}
    \hline Qubit Index 
        & Init 
        & T Gate Directions 
        & Finish 
        & T 
        & Infidelity
    \\($k$ in $Z^{2^{-k}} |+\rangle$) 
        & Basis 
        &  (the signs in $T_X^{\pm 1}, T_Z^{\pm 1}, T_X^{\pm 1}, T_Z^{\pm 1}, \dots$) 
        & Clifford 
        & Count 
        &
    
    \\\hline 
        $0$ 
        & $R_X$ 
        & \texttt{} 
        & Z 
        & 0 
        & 0
    \\\hline 
        $1$ 
        & $R_Y$ 
        & \texttt{} 
        &  
        & 0 
        & 0
    \\\hline 
        $2$ 
        & $R_Z$ 
        & \texttt{+} 
        & H 
        & 1 
        & 0
    \\\hline 
        $3$ 
        & $R_Z$ 
        & \texttt{+{}-{}-{}+{}+{}-{}+{}+{}-{}-{}+{}-{}+{}+{}+{}-{}+{}-{}-{}-{}-{}+} 
        & H,Y 
        & 22 
        & 5.8e-7
    \\\hline 
        $4$ 
        & $R_Z$ 
        & \texttt{+{}+{}+{}+{}-{}-{}-{}-{}-{}+{}-{}-{}+{}-{}-{}+{}-{}+{}+{}-{}+{}+} 
        & H 
        & 22 
        & 5.7e-7
    \\\hline 
        $5$ 
        & $R_Z$ 
        & \texttt{+{}+{}-{}-{}-{}-{}+{}+{}-{}+{}+{}-{}+{}+{}+{}+{}+{}+{}-{}-{}-} 
        & X 
        & 21 
        & 2.8e-8
    \\\hline 
        $6$ 
        & $R_Z$ 
        & \texttt{-{}-{}-{}+{}+{}-{}+{}+{}+{}+{}-{}-{}-{}+{}+{}+{}-{}-{}-{}-{}-} 
        &  
        & 21 
        & 3.2e-8
    \\\hline 
        $7$ 
        & $R_Z$ 
        & \texttt{+{}-{}+{}+{}-{}-{}+{}+{}-{}-{}-{}-} 
        & H 
        & 12 
        & 5.0e-7
    \\\hline 
        $8$ 
        & $R_Z$ 
        & \texttt{+{}+{}-{}-{}+{}+{}-{}-{}-{}+{}+{}-{}+{}+{}-{}+{}-{}-{}+} 
        & X 
        & 19 
        & 5.3e-8
    \\\hline 
        $9$ 
        & $R_Z$ 
        & \texttt{+{}-{}-{}+{}-{}-{}+{}+{}-{}+{}-{}+{}+{}-{}-{}-{}+{}-{}-{}-{}-} 
        &  
        & 21 
        & 7.2e-7
    \\\hline 
        $10$ 
        & $R_Z$ 
        & \texttt{+{}+{}-{}-{}-{}-{}+{}+{}+{}+{}+{}+{}+{}-{}-{}-{}-{}+{}+{}-} 
        & H 
        & 20 
        & 1.2e-7
    \\\hline 
        $11$ 
        & $R_X$ 
        & \texttt{} 
        &  
        & 0 
        & 5.9e-7
    \\\hline 
        $12$ 
        & $R_X$ 
        & \texttt{} 
        &  
        & 0 
        & 1.5e-7
    \\\hline\hline 
        Totals 
        &  
        &  
        &  
        & 159 
        & 3.4e-6

    \\\hline
    \end{tabular}}

%% file: assets/man/tbl-gradient-decomposition-accurate.tex
    \resizebox{\linewidth}{!}{
    \begin{tabular}{|c|c|l|c|c|c|}
    \hline Qubit Index 
        & Init 
        & T Gate Directions 
        & Finish 
        & T 
        & Infidelity
    \\($k$ in $Z^{2^{-k}} |+\rangle$) 
        & Basis 
        &  (the signs in $T_X^{\pm 1}, T_Z^{\pm 1}, T_X^{\pm 1}, T_Z^{\pm 1}, \dots$) 
        & Clifford 
        & Count 
        &
    
    \\\hline 
        $0$ 
        & $R_X$ 
        & \texttt{} 
        & Z 
        & 0 
        & 0
    \\\hline 
        $1$ 
        & $R_Y$ 
        & \texttt{} 
        &  
        & 0 
        & 0
    \\\hline 
        $2$ 
        & $R_Z$ 
        & \texttt{+} 
        & H 
        & 1 
        & 0
    \\\hline 
        $3$ 
        & $R_Z$ 
        & \texttt{+{}-{}+{}-{}-{}-{}-{}+{}+{}-{}-{}-{}-{}-{}-{}+{}+{}-{}+{}-{}-{}-{}-{}+{}-{}-{}-{}-{}-{}+{}+{}-{}+{}+{}+{}-{}-{}+{}+{}+{}-{}-{}-{}-{}+{}+{}+{}-{}+{}+} 
        & H,Y 
        & 50 
        & 0
    \\\hline 
        $4$ 
        & $R_Z$ 
        & \texttt{-{}-{}-{}-{}+{}+{}+{}-{}+{}-{}+{}-{}-{}-{}+{}+{}+{}-{}+{}+{}-{}-{}+{}-{}+{}-{}+{}-{}+{}+{}+{}+{}-{}-{}+{}+{}+{}-{}+{}-{}+{}+{}-{}-{}+{}-} 
        & H,Z 
        & 46 
        & 4.1e-16
    \\\hline 
        $5$ 
        & $R_Z$ 
        & \texttt{+{}-{}+{}-{}-{}+{}-{}-{}-{}-{}+{}-{}-{}-{}+{}+{}-{}+{}+{}+{}+{}+{}-{}-{}+{}+{}+{}-{}+{}+{}-{}-{}-{}+{}+{}-{}+{}+{}-{}-{}-{}-{}+{}-{}+{}+{}-{}+} 
        & S,H 
        & 48 
        & 6.8e-16
    \\\hline 
        $6$ 
        & $R_Z$ 
        & \texttt{-{}-{}-{}+{}+{}+{}-{}-{}-{}-{}-{}-{}-{}+{}+{}-{}-{}+{}+{}+{}-{}+{}-{}-{}+{}+{}-{}+{}+{}-{}-{}+{}-{}+{}-{}+{}+{}+{}+{}+{}+{}+{}-{}-{}-{}-{}+{}-{}+} 
        & X 
        & 49 
        & 4.1e-17
    \\\hline 
        $7$ 
        & $R_Z$ 
        & \texttt{+{}+{}-{}-{}-{}+{}+{}+{}+{}+{}-{}-{}+{}-{}-{}-{}+{}+{}+{}+{}-{}+{}-{}-{}-{}+{}-{}+{}+{}-{}+{}-{}+{}+{}+{}-{}+{}+{}+{}-{}-{}+{}+{}-{}-{}-{}-{}-{}-{}+} 
        & H 
        & 50 
        & 6.9e-16
    \\\hline 
        $8$ 
        & $R_Z$ 
        & \texttt{+{}+{}-{}+{}+{}-{}+{}+{}+{}-{}-{}+{}+{}-{}+{}+{}-{}+{}+{}+{}+{}-{}+{}+{}-{}-{}+{}+{}+{}+{}+{}+{}-{}-{}-{}-{}-{}-{}-{}-{}-{}-{}+{}-{}-} 
        & Y 
        & 45 
        & 3.0e-16
    \\\hline 
        $9$ 
        & $R_Z$ 
        & \texttt{+{}-{}-{}-{}+{}-{}-{}+{}+{}+{}+{}+{}+{}+{}+{}-{}-{}+{}-{}-{}+{}+{}-{}-{}+{}+{}-{}+{}-{}+{}+{}-{}-{}-{}+{}-{}+{}+{}-{}-{}-{}-{}+{}-{}+{}-{}+{}+} 
        & H 
        & 48 
        & 2.3e-16
    \\\hline 
        $10$ 
        & $R_Z$ 
        & \texttt{+{}+{}+{}-{}-{}+{}-{}-{}+{}-{}+{}+{}+{}+{}+{}+{}+{}-{}+{}+{}+{}+{}-{}+{}-{}+{}+{}-{}+{}-{}+{}+{}+{}-{}+{}+{}-{}-{}-{}+{}+{}+{}+{}+{}-{}+{}-{}-{}-} 
        &  
        & 49 
        & 5.0e-17
    \\\hline 
        $11$ 
        & $R_Z$ 
        & \texttt{+{}+{}-{}+{}-{}-{}+{}+{}-{}+{}+{}-{}-{}+{}+{}-{}-{}-{}-{}-{}+{}-{}+{}-{}+{}-{}+{}-{}-{}-{}+{}+{}+{}-{}-{}-{}-{}-{}-{}-{}+{}-{}+{}-{}-{}+{}+} 
        & Z 
        & 47 
        & 9.7e-16
    \\\hline 
        $12$ 
        & $R_Z$ 
        & \texttt{+{}+{}-{}-{}-{}-{}-{}-{}+{}+{}+{}+{}+{}+{}-{}-{}+{}-{}-{}+{}-{}+{}+{}-{}+{}-{}-{}-{}-{}+{}-{}+{}+{}+{}+{}+{}-{}-{}+{}+{}+{}-{}-{}-{}+} 
        & Y 
        & 45 
        & 6.8e-16
    \\\hline 
        $13$ 
        & $R_Z$ 
        & \texttt{+{}+{}+{}-{}-{}-{}+{}-{}-{}+{}+{}+{}+{}-{}+{}+{}-{}-{}+{}+{}+{}-{}-{}+{}+{}-{}-{}+{}-{}+{}+{}-{}+{}+{}+{}+{}+{}-{}-{}+{}-{}+{}+{}-{}-{}+} 
        & H,Z 
        & 46 
        & 6.9e-16
    \\\hline 
        $14$ 
        & $R_Z$ 
        & \texttt{+{}+{}+{}-{}-{}+{}-{}-{}-{}-{}-{}+{}+{}+{}-{}-{}-{}+{}+{}+{}+{}-{}+{}-{}+{}+{}+{}+{}+{}-{}+{}+{}+{}-{}-{}-{}-{}+{}+{}+{}-{}-{}-{}+{}-{}+{}-{}-{}+} 
        &  
        & 49 
        & 3.9e-16
    \\\hline 
        $15$ 
        & $R_Z$ 
        & \texttt{+{}-{}-{}-{}+{}+{}-{}+{}-{}-{}+{}-{}+{}+{}-{}-{}-{}+{}-{}-{}+{}+{}+{}-{}+{}+{}-{}+{}+{}+{}+{}-{}-{}+{}+{}+{}+{}-{}+{}+{}+{}-{}+{}-{}-{}+} 
        & H,Y 
        & 46 
        & 6.2e-16
    \\\hline 
        $16$ 
        & $R_Z$ 
        & \texttt{+{}-{}-{}+{}-{}+{}-{}+{}+{}-{}-{}+{}+{}+{}+{}+{}-{}+{}-{}+{}+{}-{}-{}-{}+{}+{}+{}+{}+{}-{}+{}-{}+{}+{}+{}+{}-{}-{}+{}-{}-{}-{}+{}-{}+{}+{}-{}+} 
        & H,Y 
        & 48 
        & 4.9e-16
    \\\hline 
        $17$ 
        & $R_Z$ 
        & \texttt{+{}-{}+{}+{}-{}-{}-{}-{}+{}-{}-{}-{}+{}+{}+{}+{}-{}+{}-{}+{}+{}+{}-{}+{}-{}+{}+{}+{}+{}-{}+{}-{}+{}-{}-{}-{}-{}+{}-{}+{}+{}-{}-} 
        &  
        & 43 
        & 5.3e-16
    \\\hline 
        $18$ 
        & $R_Z$ 
        & \texttt{+{}+{}-{}-{}-{}+{}-{}-{}+{}+{}-{}-{}+{}-{}-{}-{}+{}-{}-{}+{}-{}+{}-{}-{}-{}+{}-{}-{}-{}-{}-{}-{}-{}-{}-{}+{}+{}+{}+{}+{}+{}+{}+{}-{}-{}+{}+{}-{}-} 
        & X 
        & 49 
        & 1.8e-16
    \\\hline 
        $19$ 
        & $R_Z$ 
        & \texttt{+{}-{}-{}+{}-{}-{}+{}-{}+{}-{}+{}-{}-{}-{}-{}-{}-{}-{}+{}-{}-{}-{}+{}-{}-{}+{}-{}-{}-{}+{}+{}-{}+{}-{}-{}+{}-{}+{}+{}+{}-{}+{}-{}+{}+{}+{}-{}+{}+} 
        & Z 
        & 49 
        & 1.3e-16
    \\\hline 
        $20$ 
        & $R_Z$ 
        & \texttt{-{}-{}-{}-{}+{}-{}+{}+{}+{}+{}-{}-{}+{}-{}+{}+{}-{}-{}+{}+{}+{}-{}-{}+{}+{}+{}-{}-{}+{}+{}-{}-{}-{}-{}-{}-{}-{}+{}-{}+{}-{}+{}-{}-{}-{}+{}+{}-{}-} 
        &  
        & 49 
        & 9.7e-16
    \\\hline 
        $21$ 
        & $R_Z$ 
        & \texttt{+{}+{}-{}-{}-{}+{}+{}-{}+{}+{}-{}+{}-{}-{}+{}+{}-{}-{}-{}-{}+{}-{}+{}-{}+{}-{}-{}+{}-{}+{}+{}+{}+{}-{}-{}+{}+{}-{}+{}+{}-{}-{}-{}+{}+{}+{}+{}+} 
        & H 
        & 48 
        & 6.1e-16
    \\\hline 
        $22$ 
        & $R_Z$ 
        & \texttt{+{}-{}+{}+{}-{}+{}-{}+{}-{}-{}-{}-{}+{}+{}+{}-{}-{}+{}-{}+{}+{}-{}-{}+{}-{}+{}-{}+{}-{}+{}+{}-{}+{}+{}-{}-{}-{}+{}+{}-{}+{}-{}-{}+{}+{}+{}-{}+{}-{}-{}+{}+} 
        & H,Z 
        & 52 
        & 3.4e-16
    \\\hline 
        $23$ 
        & $R_Z$ 
        & \texttt{+{}+{}+{}-{}-{}-{}+{}-{}-{}-{}+{}-{}+{}-{}-{}+{}+{}-{}-{}+{}-{}-{}-{}-{}-{}-{}+{}-{}-{}+{}+{}-{}-{}+{}-{}+{}-{}-{}-{}+{}-{}-{}-{}+{}+{}+{}-} 
        & X 
        & 47 
        & 5.9e-16
    \\\hline 
        $24$ 
        & $R_Z$ 
        & \texttt{+{}+{}+{}+{}-{}-{}-{}-{}-{}+{}-{}+{}+{}-{}-{}-{}+{}-{}+{}+{}+{}-{}-{}-{}+{}-{}+{}-{}+{}+{}+{}+{}-{}+{}+{}+{}-{}+{}+{}+{}+{}-{}+{}+{}+{}+{}-{}-{}+} 
        & X 
        & 49 
        & 2.5e-16
    \\\hline 
        $25$ 
        & $R_Z$ 
        & \texttt{+{}-{}+{}-{}+{}+{}-{}-{}-{}+{}+{}+{}-{}+{}+{}+{}+{}-{}+{}-{}+{}-{}+{}+{}+{}+{}-{}+{}-{}+{}-{}+{}+{}+{}+{}-{}+{}+{}+{}-{}-{}-{}+{}+{}-{}+{}-{}+{}+} 
        &  
        & 49 
        & 2.9e-16
    \\\hline 
        $26$ 
        & $R_X$ 
        & \texttt{} 
        &  
        & 0 
        & 5.5e-16
    \\\hline 
        $27$ 
        & $R_X$ 
        & \texttt{} 
        &  
        & 0 
        & 1.4e-16
    \\\hline\hline 
        Totals 
        &  
        &  
        &  
        & 1102 
        & 1.1e-14

    \\\hline
    \end{tabular}}